\documentclass[reprint,amsmath,amssymb,aps,prl,superscriptaddress]{revtex4-1}

\usepackage{graphicx}% Include figure files
\usepackage{dcolumn}% Align table columns on decimal point
\usepackage{bm}% bold math
\usepackage{hyperref}% add hypertext capabilities
\usepackage{lineno}
\usepackage{multirow}
\usepackage{enumitem}
\usepackage[english]{babel}

\usepackage{xcolor}
\begin{document}

\title{Search for sterile neutrinos in MINOS and MINOS+ using a two-detector fit\vspace{-0.18cm}}

\newcommand{\Berkeley}{Lawrence Berkeley National Laboratory, Berkeley, California, 94720 USA}
\newcommand{\Cambridge}{Cavendish Laboratory, University of Cambridge, %Madingley Road, 
Cambridge CB3 0HE, United Kingdom}
\newcommand{\Cincinnati}{Department of Physics, University of Cincinnati, Cincinnati, Ohio 45221, USA}
\newcommand{\FNAL}{Fermi National Accelerator Laboratory, Batavia, Illinois 60510, USA}
\newcommand{\RAL}{Rutherford Appleton Laboratory, Science and Technology Facilities Council, Didcot, OX11 0QX, United Kingdom}
\newcommand{\UCL}{Department of Physics and Astronomy, University College London, 
%Gower Street, 
London WC1E 6BT, United Kingdom}
\newcommand{\CERN}{CERN, CH-1211 Geneva 23, Switzerland}
\newcommand{\Caltech}{Lauritsen Laboratory, California Institute of Technology, Pasadena, California 91125, USA}
\newcommand{\Alabama}{Department of Physics and Astronomy, University of Alabama, Tuscaloosa, Alabama 35487, USA}
\newcommand{\ANL}{Argonne National Laboratory, Argonne, Illinois 60439, USA}
\newcommand{\Athens}{Department of Physics, University of Athens, GR-15771 Athens, Greece}
\newcommand{\NTUAthens}{Department of Physics, National Tech. University of Athens, GR-15780 Athens, Greece}
\newcommand{\Benedictine}{Physics Department, Benedictine University, Lisle, Illinois 60532, USA}
\newcommand{\BNL}{Brookhaven National Laboratory, Upton, New York 11973, USA}
\newcommand{\CdF}{APC -- Universit\'{e} Paris 7 Denis Diderot, 10, rue Alice Domon et L\'{e}onie Duquet, F-75205 Paris Cedex 13, France}
\newcommand{\Cleveland}{Cleveland Clinic, Cleveland, Ohio 44195, USA}
\newcommand{\Dallas}{Department of Physics, University of Dallas, Irving, Texas 75062, USA}
\newcommand{\Delhi}{Department of Physics \& Astrophysics, University of Delhi, Delhi 110007, India}
\newcommand{\GEHealth}{GE Healthcare, Florence South Carolina 29501, USA}
\newcommand{\Harvard}{Department of Physics, Harvard University, Cambridge, Massachusetts 02138, USA}
\newcommand{\HolyCross}{Holy Cross College, Notre Dame, Indiana 46556, USA}
\newcommand{\Houston}{Department of Physics, University of Houston, Houston, Texas 77204, USA}
\newcommand{\IIT}{Department of Physics, Illinois Institute of Technology, Chicago, Illinois 60616, USA}
\newcommand{\Iowa}{Department of Physics and Astronomy, Iowa State University, Ames, Iowa 50011 USA}
\newcommand{\Indiana}{Indiana University, Bloomington, Indiana 47405, USA}
\newcommand{\ITEP}{High Energy Experimental Physics Department, ITEP, B. Cheremushkinskaya, 25, 117218 Moscow, Russia}
\newcommand{\JMU}{Physics Department, James Madison University, Harrisonburg, Virginia 22807, USA}
\newcommand{\LASL}{Nuclear Nonproliferation Division, Threat Reduction Directorate, Los Alamos National Laboratory, Los Alamos, New Mexico 87545, USA}
\newcommand{\Lebedev}{Nuclear Physics Department, Lebedev Physical Institute, Leninsky Prospect 53, 119991 Moscow, Russia}
\newcommand{\Lancaster}{Lancaster University, Lancaster, LA1 4YB, UK}
\newcommand{\LLL}{Lawrence Livermore National Laboratory, Livermore, California 94550, USA}
\newcommand{\LosAlamos}{Los Alamos National Laboratory, Los Alamos, New Mexico 87545, USA}
\newcommand{\Manchester}{School of Physics and Astronomy, University of Manchester, 
%Oxford Road, 
Manchester M13 9PL, United Kingdom}
\newcommand{\MIT}{Lincoln Laboratory, Massachusetts Institute of Technology, Lexington, Massachusetts 02420, USA}
\newcommand{\Minnesota}{University of Minnesota, Minneapolis, Minnesota 55455, USA}
\newcommand{\Crookston}{Math, Science and Technology Department, University of Minnesota -- Crookston, Crookston, Minnesota 56716, USA}
\newcommand{\Duluth}{Department of Physics, University of Minnesota Duluth, Duluth, Minnesota 55812, USA}
\newcommand{\Ohio}{Center for Cosmology and Astro Particle Physics, Ohio State University, Columbus, Ohio 43210 USA}
\newcommand{\Otterbein}{Otterbein University, Westerville, Ohio 43081, USA}
\newcommand{\Oxford}{Subdepartment of Particle Physics, University of Oxford, Oxford OX1 3RH, United Kingdom}
\newcommand{\PennState}{Department of Physics, Pennsylvania State University, State College, Pennsylvania 16802, USA}
\newcommand{\PennU}{Department of Physics and Astronomy, University of Pennsylvania, Philadelphia, Pennsylvania 19104, USA}
\newcommand{\Pittsburgh}{Department of Physics and Astronomy, University of Pittsburgh, Pittsburgh, Pennsylvania 15260, USA}
\newcommand{\IHEP}{Institute for High Energy Physics, Protvino, Moscow Region RU-140284, Russia}
\newcommand{\Rochester}{Department of Physics and Astronomy, University of Rochester, New York 14627 USA}
\newcommand{\RoyalH}{Physics Department, Royal Holloway, University of London, Egham, Surrey, TW20 0EX, United Kingdom}
\newcommand{\Carolina}{Department of Physics and Astronomy, University of South Carolina, Columbia, South Carolina 29208, USA}
\newcommand{\SDakota}{South Dakota School of Mines and Technology, Rapid City, South Dakota 57701, USA}
\newcommand{\SLAC}{Stanford Linear Accelerator Center, Stanford, California 94309, USA}
\newcommand{\Stanford}{Department of Physics, Stanford University, Stanford, California 94305, USA}
\newcommand{\StJohnFisher}{Physics Department, St. John Fisher College, Rochester, New York 14618 USA}
\newcommand{\Sussex}{Department of Physics and Astronomy, University of Sussex, Falmer, Brighton BN1 9QH, United Kingdom}
\newcommand{\TexasAM}{Physics Department, Texas A\&M University, College Station, Texas 77843, USA}
\newcommand{\Texas}{Department of Physics, University of Texas at Austin, 
%1 University Station C1600, 
Austin, Texas 78712, USA}
\newcommand{\TechX}{Tech-X Corporation, Boulder, Colorado 80303, USA}
\newcommand{\Tufts}{Physics Department, Tufts University, Medford, Massachusetts 02155, USA}
\newcommand{\UNICAMP}{Universidade Estadual de Campinas, IFGW, CP 6165, 13083-970, Campinas, SP, Brazil}
\newcommand{\UFG}{Instituto de F\'{i}sica, 
Universidade Federal de Goi\'{a}s, 74690-900, Goi\^{a}nia, GO, Brazil}
\newcommand{\USP}{Instituto de F\'{i}sica, Universidade de S\~{a}o Paulo,  CP 66318, 05315-970, S\~{a}o Paulo, SP, Brazil}
\newcommand{\Warsaw}{Department of Physics, University of Warsaw, %Pasteura 5, 
PL-02-093 Warsaw, Poland}
\newcommand{\Washington}{Physics Department, Western Washington University, Bellingham, Washington 98225, USA}
\newcommand{\WandM}{Department of Physics, College of William \& Mary, Williamsburg, Virginia 23187, USA}
\newcommand{\Wisconsin}{Physics Department, University of Wisconsin, Madison, Wisconsin 53706, USA}
\newcommand{\deceased}{Deceased.}

\affiliation{\ANL}
\affiliation{\Athens}
\affiliation{\BNL}
\affiliation{\Caltech}
\affiliation{\Cambridge}
\affiliation{\UNICAMP}
\affiliation{\Cincinnati}
\affiliation{\Dallas}
\affiliation{\FNAL}
\affiliation{\UFG}
\affiliation{\Harvard}
\affiliation{\HolyCross}
\affiliation{\Houston}
\affiliation{\IIT}
\affiliation{\Indiana}
\affiliation{\Iowa}
\affiliation{\Lancaster}
\affiliation{\UCL}
\affiliation{\Manchester}
\affiliation{\Minnesota}
\affiliation{\Duluth}
\affiliation{\Otterbein}
\affiliation{\Oxford}
\affiliation{\Pittsburgh}
\affiliation{\RAL}
\affiliation{\USP}
\affiliation{\Carolina}
\affiliation{\Stanford}
\affiliation{\Sussex}
\affiliation{\TexasAM}
\affiliation{\Texas}
\affiliation{\Tufts}
\affiliation{\Warsaw}
\affiliation{\WandM}

\author{P.~Adamson}
\affiliation{\FNAL}

\author{I.~Anghel}
\affiliation{\Iowa}
\affiliation{\ANL}

\author{A.~Aurisano}
\affiliation{\Cincinnati}

\author{G.~Barr}
\affiliation{\Oxford}

\author{M.~Bishai}
\affiliation{\BNL}

\author{A.~Blake}
\affiliation{\Cambridge}
\affiliation{\Lancaster}

\author{G.~J.~Bock}
\affiliation{\FNAL}

\author{D.~Bogert}
\affiliation{\FNAL}

\author{S.~V.~Cao}
\affiliation{\Texas}

\author{T.~J.~Carroll}
\affiliation{\Texas}

% ...

\collaboration{The MINOS+ Collaboration}
\noaffiliation

\begin{abstract}
A search for mixing between active neutrinos and light sterile neutrinos has been performed by looking for muon neutrino disappearance in two detectors at baselines of 1.04\,km and 735\,km, using a combined MINOS and MINOS+ exposure of $16.36\times10^{20}$ protons-on-target. A simultaneous fit to the charged-current muon neutrino and neutral-current neutrino energy spectra in the two detectors yields no evidence for sterile neutrino mixing using a 3+1 model. The most stringent limit to date is set on the mixing parameter $\sin^2\theta_{24}$ for most values of the sterile neutrino mass-splitting $\Delta m^2_{41} > 10^{-4}\,$eV$^2$.
\end{abstract}

\maketitle
The three-flavor paradigm of neutrino oscillations has been well established through the study of neutrinos produced by accelerators, nuclear reactors, the Sun, and in the atmosphere~\cite{oscSuperK,oscSNO,oscKamland,oscK2K,oscDayaBay,MINOSFinal3Flav}. It is consistent with LEP measurements of the invisible part of the decay width of the $Z$ boson that strongly constrain the number of neutrinos with $m_\nu < \frac{1}{2}m_Z$ to three~\cite{LEP2006}. Neutrino oscillations arise from the quantum mechanical interference between the neutrino mass states as they propagate. These mass states are related to the weak interaction flavor eigenstates by the PMNS mixing matrix~\cite{Pontecorvo1967,Pontecorvo1968,MNS1962}. This unitary $3\times3$ matrix is commonly parameterized in terms of three mixing angles, $\theta_{12}$, $\theta_{13}$ and $\theta_{23}$, and a CP-violating phase $\delta_{CP}$. The frequencies of the oscillations are given by the differences between the squares of the masses (mass-splittings), $\Delta m^{2}_{kj} \equiv m^{2}_{k} - m^{2}_{j}$. These are $\Delta m^{2}_{21}$, $\Delta m^{2}_{31}$ and $\Delta m^{2}_{32}$, of which only two are independent. However, some experimental results are in tension with the three-flavor paradigm: anomalous appearance of $\bar{\nu}_e$ in short-baseline $\bar{\nu}_\mu$ beams at LSND~\cite{LSNDSterile} and MiniBooNE~\cite{newMiniBooNE}; depletion of $\nu_e$ with respect to predicted rates from radioactive calibration sources in gallium experiments~\cite{GalliumSummary}; and $\bar{\nu}_e$ rate deficits seen in reactor neutrino experiments with respect to recent reactor flux calculations~\cite{ReactorSummary}, though this anomaly has been weakened by Daya Bay's reactor fuel cycle measurements~\cite{dayaBayFlux}, and by observations of spectral distortions not predicted by flux calculations~\cite{PhysRevLett.118.042502}. These data can be accommodated by a fourth neutrino state at a mass-splitting scale of approximately~$1\,$eV$^2$. This new state must not couple through the weak interaction and is thus referred to as \emph{sterile}. The MINOS and MINOS+ long-baseline neutrino experiments are sensitive to oscillations involving sterile neutrinos. Following the previous searches reported by MINOS~\cite{MINOSSterile2016,MINOSDayaBay2016}, this Letter reports results of a significantly higher sensitivity search for sterile neutrinos using an improved analysis method and incorporating data collected by the MINOS+ Experiment.

A simple model of neutrino flavor mixing that incorporates a sterile neutrino is the 3+1 model, whereby a new flavor state $\nu_s$ and a new mass state $\nu_4$ are added to the existing three-flavor formalism. In this model, the extended PMNS matrix is a $4\times4$ unitary matrix, which introduces three additional mixing angles $\theta_{14}$, $\theta_{24}$ and $\theta_{34}$, as well as two CP-violating phases, $\delta_{14}$ and $\delta_{24}$, in addition to $\delta_{13}\equiv\delta_{\textrm{CP}}$. Three new mass-splitting terms can be defined and in this analysis results are expressed as a function of the $\Delta m^{2}_{41}$ mass-splitting.

While sterile neutrinos can help to accommodate some observed data, other experimental searches have reported null results. The MINOS Collaboration have published results~\cite{MINOSSterile2016} from a sterile neutrino search using an exposure of $10.56\times10^{20}$ protons-on-target (POT) from the NuMI $\nu_\mu$ beam~\cite{NuMIBeam2016} with a $3\,$GeV peak beam neutrino energy. A joint analysis with Daya Bay and Bugey-3 constrained anomalous $\nu_\mu$ to $\nu_e$ transitions~\cite{MINOSDayaBay2016}. A search for anomalous atmospheric neutrino oscillations by IceCube set limits on part of the sterile neutrino parameter space~\cite{IceCubeSterile,IceCubeDC}.

This Letter presents results using an additional exposure of $5.80\times10^{20}$ POT from MINOS+, collected in the same detectors as MINOS with a $\nu_\mu$ energy distribution peaked at $7\,$GeV, well above the $1.6\,$GeV energy corresponding to the maximum three-flavor disappearance oscillation probability at $735\,$km. 
The broader energy range covered with high statistics improves the MINOS+ sensitivity to exotic phenomena such as sterile neutrinos with respect to MINOS.
The previous analysis\cite{MINOSSterile2016} was based on the ratio between the measured neutrino energy spectra in the two detectors (Far-over-Near ratio), whereas this analysis employs a two-detector fit method, directly fitting the reconstructed neutrino energy spectra in the two detectors to significantly improve the sterile neutrino sensitivity for $\Delta m^{2}_{41} > 10\,$eV$^2$. 

The MINOS/MINOS+ Experiments operated two on-axis detectors, the Near Detector (ND) located at Fermilab, $1.04\,$km from the NuMI beam target, and the larger Far Detector (FD)~\cite{MINOSNIM} located $735\,$km downstream in the Soudan Underground Laboratory in Minnesota. The detectors were functionally equivalent magnetized tracking sampling calorimeters with alternating planes of scintillator strips oriented at $\pm45^{\circ}$ to the vertical, interleaved with $2.54\,$cm-thick steel planes. The beam is produced by directing 120\,GeV protons from Fermilab's Main Injector accelerator onto a graphite target and focusing the emitted $\pi$ and $K$ mesons into a 625\,m pipe where they decay into a predominantly $\nu_\mu$ beam.

The analysis presented here utilizes both the charged-current~(CC)~$\nu_\mu$ and the neutral-current~(NC) data samples from MINOS and MINOS+. The analysis uses exact oscillation probabilities, but approximations are made in the text to demonstrate the sensitivity to the sterile neutrino oscillation parameters: terms related to $\Delta m^2_{21}$ are considered to be negligible, hence $\Delta m^{2}_{32} \approx \Delta m^{2}_{31}$; and $\Delta m^{2}_{41} \gg \Delta m^{2}_{31}$ is assumed, such that $\Delta m^{2}_{41} \approx \Delta m^{2}_{42} \approx \Delta m^{2}_{43}$. The oscillation probabilities can be expanded to second order in $\sin\theta_{13}$, $\sin\theta_{14}$~\cite{theta14ZeroJustification}, $\sin\theta_{24}$~\cite{MINOSSterile2016} and $\cos2\theta_{23}$. Consequently, the $\nu_\mu$ survival probability for a neutrino that traveled a~distance $L$ with energy $E$ is:
\begin{eqnarray}
P\left(\nu_\mu\rightarrow\nu_\mu\right)\approx&1&-\sin^{2}2\theta_{23}\cos2\theta_{24}\sin^{2}\left(\frac{\Delta m^{2}_{31} L }{4E}\right) \nonumber \\
&-&\sin^{2}2\theta_{24}\sin^{2}\left(\frac{\Delta m^{2}_{41}L}{4E}\right).
\end{eqnarray}
Therefore, the CC $\nu_\mu$ disappearance channel has sensitivity to $\theta_{24}$ and $\Delta m^{2}_{41}$, in addition to the three-flavor oscillation parameters $\Delta m^{2}_{32}$ and $\theta_{23}$. Similarly, the NC survival probability is given by:  
\begin{eqnarray}
 P_{\textrm{NC}} &=& 1 - P\left(\nu_\mu\rightarrow\nu_s\right) \nonumber \\
&\approx& 1 -\cos^{4}\theta_{14}\cos^{2}\theta_{34}\sin^{2}2\theta_{24}\sin^{2}\left(\frac{\Delta m^{2}_{41}L}{4E}\right) \nonumber \\
& &- \sin^{2}\theta_{34}\sin^{2}2\theta_{23}\sin^{2}\left(\frac{\Delta m^{2}_{31} L }{4E}\right) \\
& &+\frac{1}{2}\sin\delta_{24}\sin\theta_{24}\sin2\theta_{34}\sin2\theta_{23}\sin\left(\frac{\Delta m^{2}_{31} L }{2E}\right). \nonumber
\end{eqnarray}
The NC sample has sensitivity to both $\theta_{24}$ and $\Delta m^{2}_{41}$, and further depends on $\theta_{14}$, $\theta_{34}$ and $\delta_{24}$. The sensitivity to four-flavor neutrino oscillations is weaker in the NC channel than the CC channel as a result of the poorer energy resolution due to the invisible outgoing neutrino in the final state of the NC interaction and the lower NC cross section.

The effect of a sterile neutrino would be a modulation of the neutrino energy spectra on top of the well-measured three-flavor oscillation~\cite{MINOSDis2013}. The actual effect depends strongly on the value of $\Delta m^{2}_{41}$. For values of \mbox{$\Delta m^{2}_{41} \lesssim 0.1\,$eV$^{2}$}, the sterile-driven oscillations are seen as an energy-dependent modification to the FD spectra. In the range \mbox{$0.1\lesssim \Delta m^{2}_{41} \lesssim 1\,$eV$^{2}$}, oscillations still only affect FD observations, but now they are rapid, that is, they have a wavelength comparable to or shorter than the energy resolution of the detector so are seen as a deficit in the event rate, constant in energy. For \mbox{$1\,$eV$^{2} \lesssim \Delta m^{2}_{41} \lesssim 100\,$eV$^{2}$}, oscillations occur in the ND along with rapid oscillations averaging in the FD. Finally, for values of \mbox{$\Delta m^{2}_{41} \gtrsim 100\,$eV$^{2}$}, rapid oscillations occur upstream of the ND, causing event rate deficits in both detectors.

\begin{figure}{}
	\begin{center}
    \includegraphics[scale=0.44]{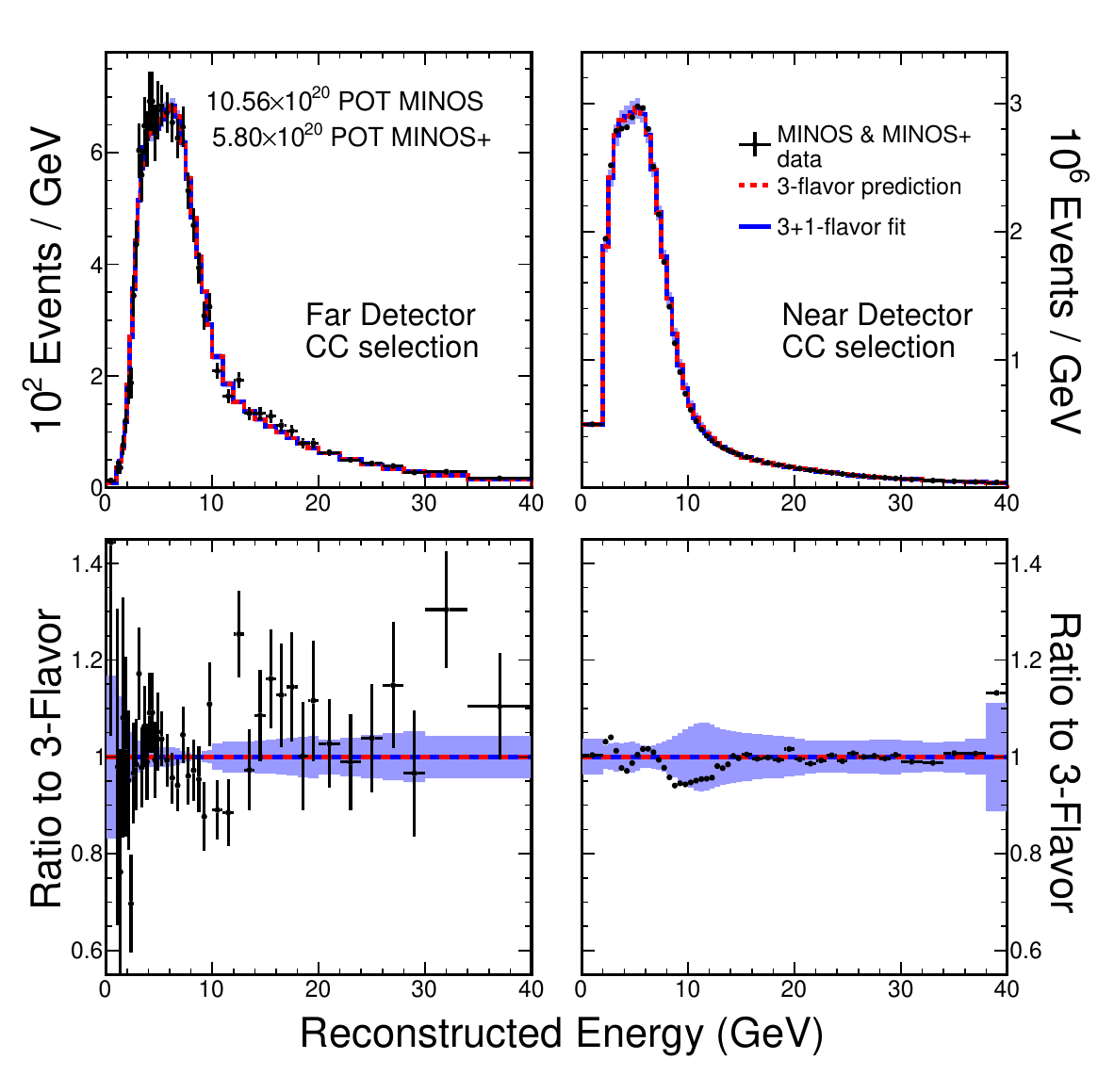}
    \end{center}
    \vspace{-20pt}
    \caption{\label{fig:CCspectra}The summed MINOS and MINOS+ CC reconstructed energy spectra for events selected in the FD (left) and ND (right). Data points in black are compared to both the three-flavor prediction (red line) and the four-flavor best-fit prediction $\left(\sin^2\theta_{24} = 1.1\times 10^{-4}, \Delta m^2_{41} = 2.325\times 10^{-3} \textrm{eV}^2 \right)$ and its systematic uncertainty (blue line and shaded region). Also shown are the ratios between data and simulation.}
\end{figure}

\begin{figure}
	\centering
    \includegraphics[scale=0.44]{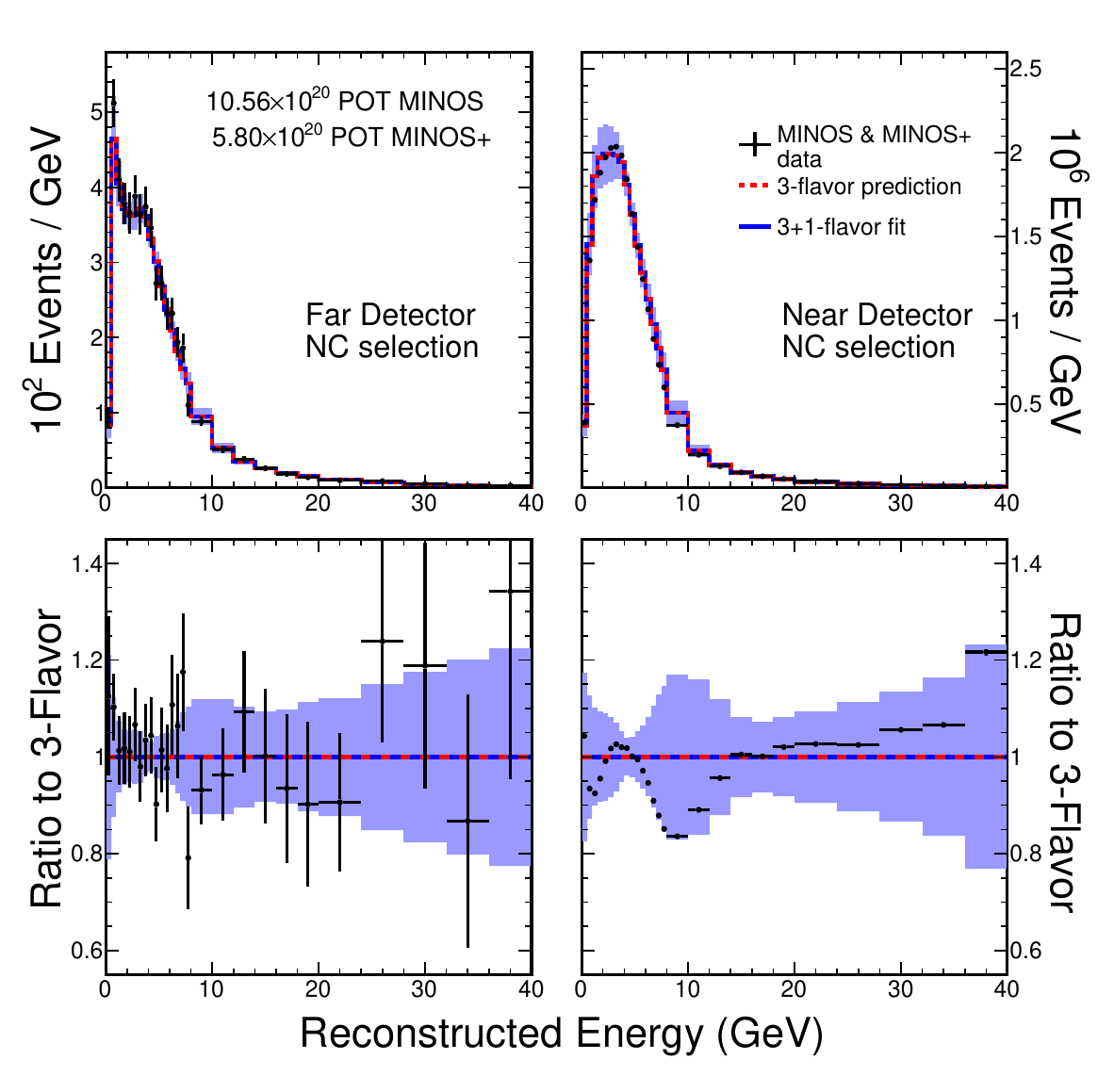}
    \caption{\label{fig:NCspectra}The summed MINOS and MINOS+ NC reconstructed energy spectra for events selected in the FD (left) and ND (right). Data points in black are compared to both the three-flavor prediction (red line) and the four-flavor best-fit prediction $\left(\sin^2\theta_{24} = 1.1\times 10^{-4}, \Delta m^2_{41} = 2.325\times 10^{-3} \textrm{eV}^2 \right)$ and its systematic uncertainty (blue line and shaded region). Also shown are the ratios between data and simulation.}
\end{figure}

For the MINOS data, the event classification algorithm remains unchanged~\cite{MINOSSterile2016}, while for MINOS+ the event selection and reconstruction were re-tuned to account for a four-fold ND occupancy increase. From Monte Carlo (MC) studies, defining the denominator of efficiency as all true NC interactions reconstructed within the detector's fiducial volume, the MINOS+ beam NC selection in the ND has an efficiency of 79.9\% and purity of 60.3\%, and in the FD, the efficiency is 86.5\% with 64.9\% purity~\cite{junting}. The beam CC selection in the ND is 56.4\% efficient with a purity of 99.1\%, and the FD CC selection has 85.1\% efficiency and 99.3\% purity~\cite{junting}. The MINOS era efficiencies and purities agree with MINOS+ within a few percent. The CC and NC reconstructed neutrino energy spectra for the ND and FD are shown in Figs.~\ref{fig:CCspectra} and~\ref{fig:NCspectra}, respectively. The three-flavor and best-fit four-flavor predictions are also shown.
Visual representation of the effects of correlated systematics uncertainties is not straightforward, so we perform a decorrelation of the systematic uncertainty covariance matrix using conditional multivariate Gaussian distributions~\cite{jacobTodd} to produce the uncertainty bands.
The decorrelation procedure is conducted through the iterative conditioning of the expected distribution of each reconstructed energy bin upon all other bins of the observed event spectrum~\footnote{See Supplemental Material for further details, which contains Refs.~\cite{Asimov,ref:fluka1,ref:fluka2,NA49Exp,MinervaExp,ref:matrixpram,CCFRExp,NOMADExp,ref:flugg,Geant4:2003,Geant4:2006,MEC,MECMiniBooNE,RPA,RPAMinerva,MinosCalDet,intranuke,HadESyst,Conway:2011in,ATLAS_Asimov,katz_uli_2018,Patterson:2007zz,JAMES1975343,CovMx}}. Accounting for the systematic correlations simultaneously decreases the effective uncertainty and improves the agreement between the predicted spectra and observed data, thereby adding confidence in the modeling of the systematic uncertainties.

The previous Far-over-Near ratio method was limited by a reduction in the sensitivity to the $\theta_{24}$ mixing angle at high values of $\Delta m^2_{41}$, where the oscillations occur upstream of the ND and cancel in the ratio. Simultaneous measurements in both detectors extends the observed range of experimental L/E and consequently yields broader sensitivity to the sterile mass-splitting. Furthermore, the uncertainty on the ratio was dominated by the FD statistical uncertainty, which limited the high statistical power of the ND in the fit. To improve the overall sensitivity and better utilize the high-statistics ND data sample, the two-detector fit method has been developed. 

The MINOS three-flavor oscillation analyses use the ND data to tune the MC flux simulation to provide an accurate flux prediction in the FD. In the context of this 3+1-flavor analysis, oscillations can occur in both detectors, and therefore the beam tuning approach~\cite{minosDisPRD} assuming no oscillations at the ND is invalid. 

The flux prediction uses a combination of the MINERvA PPFX flux~\cite{MinervaFlux2016,leoThesis} that uses only hadron production data~\footnote{The MINERvA publication~\cite{MinervaFlux2016} incorporates the $\nu_e e \rightarrow \nu_e e$ scattering rate in the published fluxes. This constraint is not used in this analysis.} and the published data of the $\pi^+/K^+$ hadron production ratio to which FLUKA is tuned~\cite{mippFluka}. For general applicability to both MINOS and MINOS+ running, which used different target designs, we chose to use the version of PPFX using thin-target hadron production data. We extract eight parameters used to warp an empirical parameterization FLUKA $\pi^+$ hadron production as a function of $p_{T}$ and $p_{Z}$ using a sample of simulated ND PPFX-weighted pion-parent interactions in configurations with the magnetic focusing horns powered off and on.  The $\pi^+/K^+$ ratio is used to extend the results of this fit to the kaon flux component. 

A search is performed simultaneously in both detectors for oscillations due to sterile neutrinos by minimizing the sum of the following $\chi^{2}$ statistic for selected candidate CC and NC events:
\begin{eqnarray}\label{eq:chiFunc}
 \chi^{2} = \sum_{i = 1}^{N} \sum_{j = 1}^{N} (x_{i}-\mu_{i})[\bm{V}^{-1}]_{ij}(x_{j}-\mu_{j}) + \textrm{penalty},
\end{eqnarray}
where the number of events observed in data and the MC prediction are denoted by $x_{i}$ and $\mu_{i}$, respectively. The index $i = 1,...,N$ labels the reconstructed energy bins from 0 to 40\,GeV in each detector with $N$ being the sum of ND and FD bins. The predicted number of events $\mu_{i}$ is varied using an MC simulation with exact forms of all oscillation probabilities in vacuum. The impact of the matter potential was found to be very small~\cite{ashleyThesis} and is neglected. In order to account for rapidly varying oscillations at short baselines, the calculation of neutrino oscillation probabilities in the ND uses the fully-simulated propagation distance from the point of meson decay to the neutrino interaction. These variations in path length are negligible in the FD, where a point source is assumed. 

The penalty term in Eq.~(\ref{eq:chiFunc}) is a weak constraint on $\Delta m^{2}_{31}$ to ensure it does not deviate too far from its measured value and become degenerate with $\Delta m^{2}_{41}$. 

The matrix $\bm{V}^{-1}$ is the inverse of the $N \times N$ covariance matrix that incorporates the sum of the statistical and systematic uncertainties:
\begin{eqnarray}\label{eq:systSum}
\bm{V} &= V_{\textrm{stat}} + V_{\textrm{scale}} + V_{\textrm{hp}} + V_{\textrm{xsec}} \nonumber \\ &+ V_{\textrm{bkgd}} + V_{\textrm{beam}} + V_{\textrm{other}}.
\end{eqnarray}
The general structure of the covariance matrices has four quadrants corresponding to the FD covariance matrix, the ND covariance matrix, and cross-term matrices encoding the covariance between the detectors. This treatment ensures consistency between the two detectors when ambiguities might otherwise exist between the shape of systematic fluctuations and neutrino oscillation signals.

$V_{\textrm{stat}}$ encodes the statistical uncertainty in each bin assuming Poisson statistics. The magnitude of this uncertainty is markedly different in the FD and ND given the difference in event rates.  In the FD, the statistical uncertainty is at most 13\% and averages approximately 7\% across all energy bins, while in the ND the statistical uncertainty is negligible. The statistical error only affects the diagonal elements of the covariance matrix.

$V_{\textrm{scale}}$ accounts for energy-scale uncertainties. For reconstructed muon tracks this is $\pm$2\% ($\pm$3\%) for energies measured by range (curvature)~\cite{minosSysts}. The hadronic energy scale uncertainty consists of $\pm$5.7\% from calibration, and further uncertainties from final-state interactions of hadrons within the nucleus~\cite{minosDisPRD}.

$V_{\textrm{hp}}$ accounts for the hadron production systematic uncertainty associated with the flux prediction. The uncertainties of each of the eight extracted parameters are used to generate the covariance matrix.

$V_{\textrm{xsec}}$ accounts for neutrino cross section systematic uncertainties~\cite{NEUGEN}. Details of the uncertainties considered for the different CC cross sections are given in Ref.~\cite{junting}. Note that all cross section systematic uncertainties are shape uncertainties with the exception of the 3.5\% total cross-section systematic uncertainty. This uncertainty level is justified even at large $\Delta m^2_{41}$, by high-energy cross section measurements at CCFR which showed no indications of deviations from a linear dependence on energy over a broad energy range~\cite{CCFRXSec}. The uncertainties considered for the NC cross sections are as follows: vary the axial mass $M_A^{QE}$ by ${+35}/{-15}$\%, $M_{A}^{RES}$ by ${+25}/{-15}$\%, the KNO scaling parameters~\cite{KNO} for multiplicities of 2 and 3 by $\pm33$\%, and a total cross-section variation of $\pm5$\% motivated by the difference between the measured and simulated NC/CC ratio of observed interactions.

$V_{\textrm{bkgd}}$ accounts for possible mismodeling of backgrounds in the selected CC and NC samples. The CC sample backgrounds are dominated by NC interactions and the NC component is varied by 30\% (20\%) for MINOS+ (MINOS). The NC sample has background contributions from $\nu_e$ and $\nu_\mu$ CC events in the ND and $\nu_e$, $\nu_\mu$ and $\nu_\tau$ CC events in the FD. The $\nu_e$ and $\nu_\tau$ components have a minimal impact, hence only the CC $\nu_\mu$ component is varied by $\pm15$\%.

$V_{\textrm{beam}}$ incorporates systematic uncertainties from the beam-line~\cite{NuMIBeam2016,beamSyst,Zarko}. It includes possible mismodeling of the horn current ($\pm 2$\%), the horn current distribution (exponential or linear), the horn position ($\pm 0.5\,$mm), the material in the horns ($\pm 5\%$ in atomic number), the beam width ($\pm 0.2\,$mm) and position ($\pm 0.5\,$mm), and the target position ($\pm 2\,$mm).

$V_{\textrm{other}}$ includes terms for a range of additional systematic uncertainties, including relative detector acceptance~\cite{junting}, relative efficiency of reconstruction ($\pm$1.6\% for CC, $\pm$2.2\% for NC~\cite{MINOSSterile2016}), removal of poorly reconstructed NC interactions~\cite{aDevan}, and POT normalization ($\pm$2.0\%). 

\begin{table}
 \begin{tabular}{l c c}
   \hline \hline \\[-0.25cm]
   Uncertainty & \multicolumn{2}{c}{Sensitivity to $\sin^{2}\theta_{24}$ at:} \\
   & $\Delta m^2_{41}$ = 1\,eV$^2$~ & ~$\Delta m^2_{41}$ = 1000\,eV$^2$ \\[0.1cm]
   \hline \\[-0.25cm]
   Statistics only & 0.0008 & 0.0002 \\
   +Energy scale & 0.0054 & 0.0003 \\
   +Hadron production & 0.0131 & 0.0063 \\
   +Cross section & 0.0138 & 0.0103 \\
   +Background & 0.0141 & 0.0112 \\
   +Beam & 0.0143 & 0.0128 \\
   +Other & 0.0153 & 0.0165 \\
%   \\
%   Combined & 100\% & 100\% \\
   \hline \hline
 \end{tabular}

 \caption{\label{tab:systEffect}The reduction in $\sin^2\theta_{24}$ exclusion sensitivity caused by accumulation of systematic sources at two values of $\Delta m^2_{41}$. The systematic uncertainty sources are given in Eq.~(\ref{eq:systSum}).}
\end{table}

The effect of each systematic uncertainty category on the sensitivity to $\sin^2\theta_{24}$ 
is shown in Table~\ref{tab:systEffect} for two sample values of $\Delta m^2_{41}$.

In the fit, the oscillation parameters $\theta_{23}$, $\theta_{24}$, $\theta_{34}$, $\Delta m^2_{31}$ and $\Delta m^2_{41}$ are allowed to float, while the other oscillation parameters are held at fixed values. The penalty term constrains $\Delta m^2_{31} = (2.5 \pm 0.5) \times 10^{-3}\,$eV$^2$~\cite{ashleyThesis}. The solar parameters are set at values of $\sin^{2}\theta_{12} = 0.307$ and $\Delta m^2_{21} = 7.54\times10^{-5}\,$eV$^2$, based on a three-flavor global fit~\cite{solarParsJustification}.

In the case where $\Delta m^2_{41}$ is sufficiently larger than $\Delta m^2_{31}$, $|U_{e4}|$ is constrained by unitarity considerations, which imply $\sin^2\theta_{14} < 0.036$ at 1$\sigma$~\cite{PhysRevD.93.113009}.  In the degenerate case, $\theta_{14}$ is constrained by measurements of $\theta_{13}$~\cite{solarParsJustification}, and $\theta_{13} = \pi/2$. A sensitivity study was performed with $\theta_{14}$ allowed to vary under these constraints, and $\delta_{13}$, $\delta_{14}$ and $\delta_{24}$ allowed to float freely. The resulting sensitivity is very similar to the case where these parameters were set to zero. This analysis hence has very minimal sensitivity to $\sin^2\theta_{14}$ under the aforementioned constraints so it is set to zero. The analysis is also approximately independent of $\delta_{13}$, $\delta_{14}$ and $\delta_{24}$, hence all three phases are set to zero.

\begin{figure}
	\centering
    \includegraphics[trim=0 10 0 0, clip,scale=0.45]{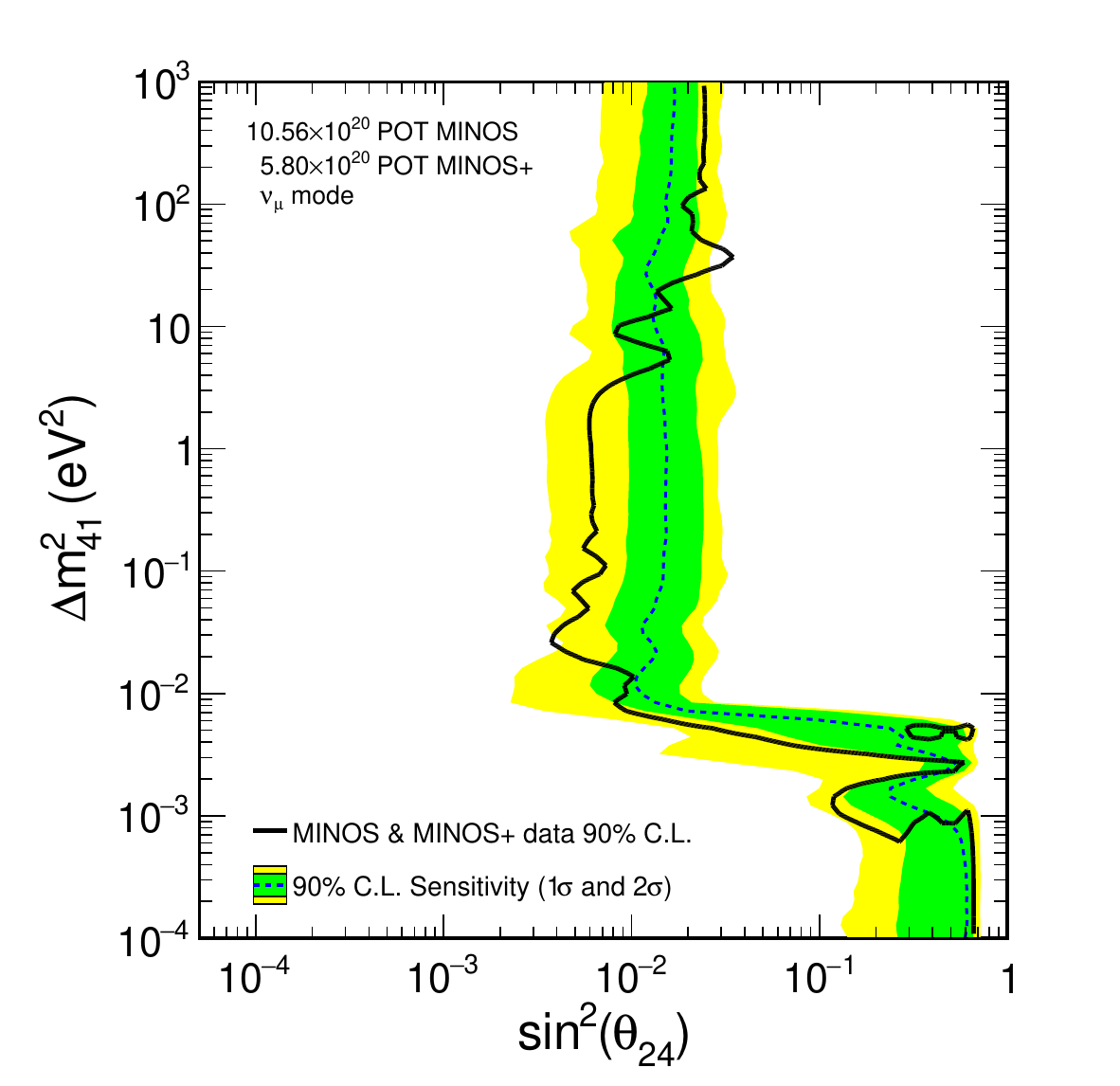}
    \caption{\label{fig:limit}The MINOS and MINOS+ 90\% Feldman-Cousins exclusion limit in the $\left( \sin^2\theta_{24},\Delta m^{2}_{41} \right)$ plane from the data (black line) compared to the expected sensitivity (blue line, $1\sigma$ green band and $2\sigma$ yellow band). For $\Delta m^{2}_{41} > 10^{3}\,$eV$^2$ the limits continue vertically upwards.}
\end{figure}

The fit proceeds by dividing the $\left( \sin^2\theta_{24},\Delta m^{2}_{41} \right)$ parameter space into fine bins ranging from $10^{-3}$ to $1$ in $\sin^2\theta_{24}$ and $10^{-4}\,$eV$^2$ to $10^{3}\,$eV$^2$ in $\Delta m^{2}_{41}$. At each point in the parameter space, the function given in Eq.~(\ref{eq:chiFunc}) is minimized with respect to the three remaining oscillation parameters $\theta_{23}$, $\theta_{34}$ and $\Delta m^2_{31}$, and the penalty term. The difference in $\chi^{2}$ at each point, compared to the global minimum $\chi^{2}_{\textrm{min}}=99.308$ (140 degrees of freedom), is shown in Fig.~\ref{fig:limit} as a $90\%$~C.L. contour interpreted using the Feldman-Cousins procedure~\cite{fcPaper}. The 3+1 model best-fit $\chi^2$ at the global minimum $\left(\sin^2\theta_{24} = 1.1\times 10^{-4}, \Delta m^2_{41} = 2.325\times 10^{-3} \textrm{eV}^2 \right)$ differs from the three-flavor model by $\Delta \chi^2 < 0.01$, and the corresponding predicted neutrino energy spectra are shown by the blue lines in Fig.~\ref{fig:CCspectra} and \ref{fig:NCspectra}. Figure~\ref{fig:limit} also shows the median sensitivity and the $1\sigma$ and $2\sigma$ sensitivity bands from a large number of pseudo-experiments generated by fluctuating the three-flavor simulation according to the 
covariance matrix $\bm{V}$ and the uncertainties on the three-flavor oscillation parameters~\cite{NuFit}.

The measured contour lies well within the $2\sigma$ sensitivity band. Fitted values of $\theta_{34}$ are found to be small across the parameter space, with the value at the best-fit point $\theta_{34} = 8.4\times10^{-3}$ , and show little correlation with $\theta_{24}$. For high $\Delta m^2_{41}$ values, where sterile oscillations produce normalization shifts at both the ND and FD, shape uncertainties are nearly irrelevant.  Therefore, the strength of the limit in this region is driven by the constraint on the total CC cross-section and unitarity constraints related to the observed near-maximal value of $\sin^2\theta_{23}$~\cite{NuFit}. At the best-fit point $\sin^2 2\theta_{23} = 0.920$.

\begin{figure}
	\centering
    \includegraphics[trim=0 10 0 0, clip,scale=0.45]{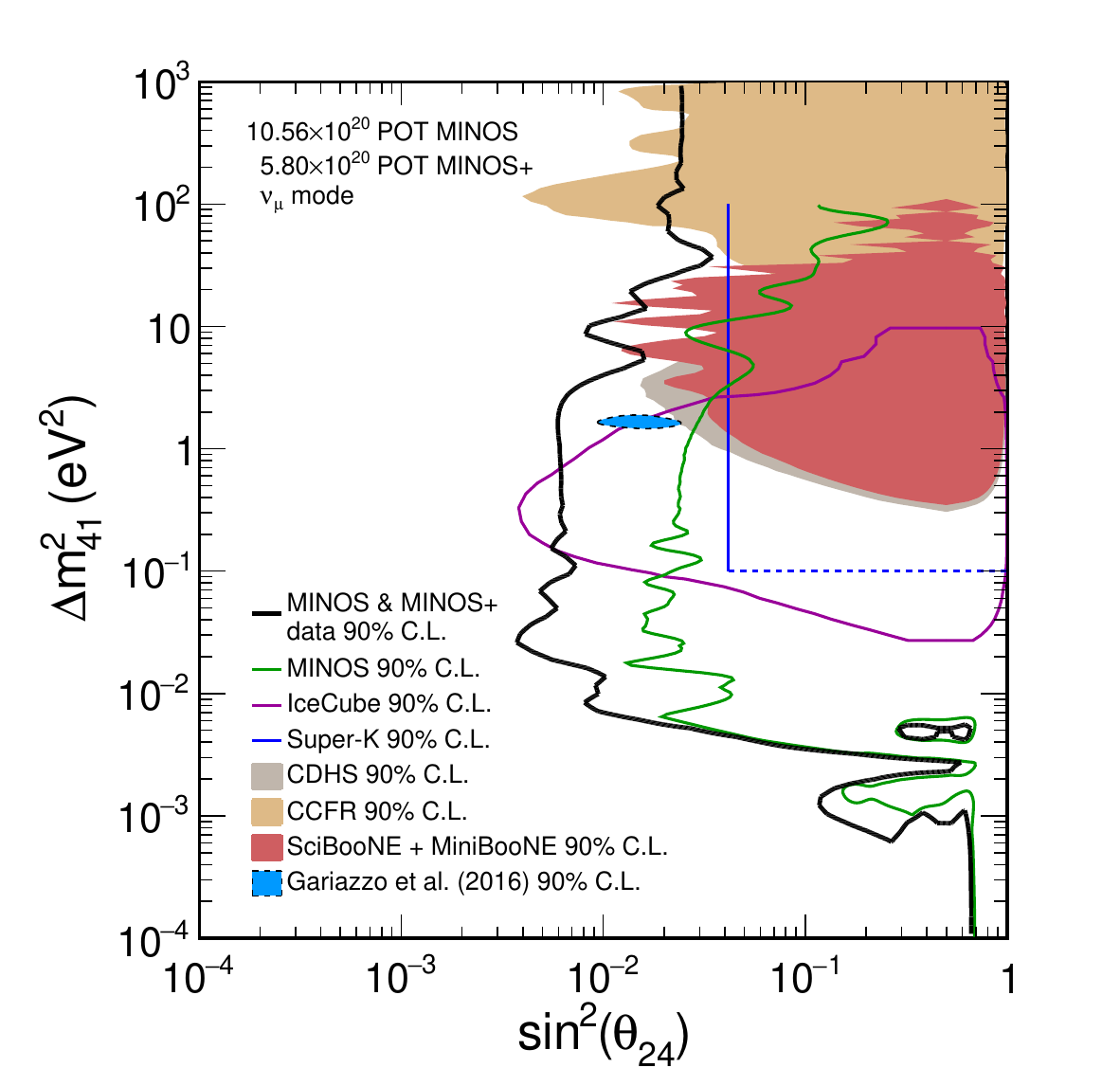}
    \caption{\label{fig:compareLimit}The MINOS and MINOS+ 90\% Feldman-Cousins exclusion limit compared to the previous MINOS result~\cite{MINOSSterile2016} and results from other experiments~\cite{IceCubeSterile,SKSterile,CDHSWSterile,CCFRSterile,SciMBSterile}. The Gariazzo \emph{et al.} region is the result of a global fit to neutrino oscillation data~\cite{GiuntiSterile}.}
\end{figure}

No evidence of mixing between active and sterile neutrinos is observed, and a stringent limit on $\theta_{24}$ is set for all values of $\Delta m^2_{41}$ above $10^{-2}\,$eV$^2$. The low sensitivity in the region $\Delta m^{2}_{41} < 10^{-2}$ eV$^{2}$ arises from degeneracies with the atmospheric mass-splitting $\Delta m^{2}_{31}$. The upper island occurs at $\Delta m^{2}_{41} = 2\Delta m^{2}_{31}$, and the dip below occurs at $\Delta m^{2}_{41} = \Delta m^{2}_{31}$. The MINOS/MINOS+ result is compared to results from other experiments in Fig.~\ref{fig:compareLimit}, showing it to be the leading limit over the majority of the range of $\Delta m^{2}_{41}$. At fixed values of $\Delta m^{2}_{41}$ the data provide limits on the mixing angles $\theta_{24}$ and $\theta_{34}$. At $\Delta m^{2}_{41} = 0.5\,$eV$^2$,  we find $\sin^{2}\theta_{24} < [0.006~(90\%$ C.L.$), 0.008~(95\%$ C.L.$)]$ and $\sin^{2}\theta_{34} < [0.41~(90\%$ C.L.$), 0.49~(95\%$ C.L.$)]$.

The MiniBooNE result~\cite{newMiniBooNE} observes a significant excess in the $\nu_{e}$ and $\bar{\nu}_{e}$ appearance channels over a short baseline. The allowed region for this result, interpreted in terms of sterile-neutrino-driven oscillations, is presented in terms of the effective mixing parameter $\sin^{2}2\theta_{\mu e}$, which can be directly compared with the MINOS/MINOS+ limit through a combination with reactor disappearance experiments~\cite{MINOSDayaBay2016}. Since $\sin^{2}2\theta_{\mu e} = \sin^{2}2\theta_{14}\sin^{2}\theta_{24}$ in the 3+1 model, it is possible to make a direct comparison via the 90\% C.L. unitarity bound of $\sin^2 2\theta_{14} < 0.18$~\cite{PhysRevD.93.113009}. MINOS/MINOS+ excludes the entire MiniBooNE 90\% C.L. allowed region at 90\% C.L. This implies a tension between the MiniBooNE and MINOS/MINOS+ results.

In conclusion, the joint analysis of data from the MINOS and MINOS+ experiments sets leading and stringent limits on mixing with sterile neutrinos in the 3+1 model for values of $\Delta m^{2}_{41} > 10^{-2}\,$eV$^2$ through the study of $\nu_\mu$ disappearance. The final year of MINOS+ data, corresponding to 40\% of the total MINOS+ exposure, combined with ongoing analysis improvements, will increase the sensitivity of future analyses even further.

This document was prepared by the MINOS/MINOS+ Collaboration using the resources of the Fermi National Accelerator Laboratory (Fermilab), a U.S. Department of Energy, Office of Science, HEP User Facility. Fermilab is managed by Fermi Research Alliance, LLC (FRA), acting under Contract No. DE-AC02-07CH11359. This work was supported by the U.S. DOE; the United Kingdom STFC; the U.S. NSF; the State and University of Minnesota; and Brazil's FAPESP, CNPq and CAPES. We thank the personnel of Fermilab's Accelerator and Scientific Computing Divisions and the crew of the Soudan Underground Laboratory for their effort and dedication. We thank the Texas Advanced Computing Center at The University of Texas at Austin for the provision of computing resources. We acknowledge fruitful cooperation with Minnesota DNR.

\bibliographystyle{apsrev4-1}
\bibliography{mainbib}

\begin{thebibliography}{73}%
\makeatletter
\providecommand \@ifxundefined [1]{%
 \@ifx{#1\undefined}
}%
\providecommand \@ifnum [1]{%
 \ifnum #1\expandafter \@firstoftwo
 \else \expandafter \@secondoftwo
 \fi
}%
\providecommand \@ifx [1]{%
 \ifx #1\expandafter \@firstoftwo
 \else \expandafter \@secondoftwo
 \fi
}%
\providecommand \natexlab [1]{#1}%
\providecommand \enquote  [1]{``#1''}%
\providecommand \bibnamefont  [1]{#1}%
\providecommand \bibfnamefont [1]{#1}%
\providecommand \citenamefont [1]{#1}%
\providecommand \href@noop [0]{\@secondoftwo}%
\providecommand \href [0]{\begingroup \@sanitize@url \@href}%
\providecommand \@href[1]{\@@startlink{#1}\@@href}%
\providecommand \@@href[1]{\endgroup#1\@@endlink}%
\providecommand \@sanitize@url [0]{\catcode `\\12\catcode `\$12\catcode
  `\&12\catcode `\#12\catcode `\^12\catcode `\_12\catcode `\%12\relax}%
\providecommand \@@startlink[1]{}%
\providecommand \@@endlink[0]{}%
\providecommand \url  [0]{\begingroup\@sanitize@url \@url }%
\providecommand \@url [1]{\endgroup\@href {#1}{\urlprefix }}%
\providecommand \urlprefix  [0]{URL }%
\providecommand \Eprint [0]{\href }%
\providecommand \doibase [0]{http://dx.doi.org/}%
\providecommand \selectlanguage [0]{\@gobble}%
\providecommand \bibinfo  [0]{\@secondoftwo}%
\providecommand \bibfield  [0]{\@secondoftwo}%
\providecommand \translation [1]{[#1]}%
\providecommand \BibitemOpen [0]{}%
\providecommand \bibitemStop [0]{}%
\providecommand \bibitemNoStop [0]{.\EOS\space}%
\providecommand \EOS [0]{\spacefactor3000\relax}%
\providecommand \BibitemShut  [1]{\csname bibitem#1\endcsname}%
\let\auto@bib@innerbib\@empty
%</preamble>
\bibitem [{\citenamefont {Fukuda}\ \emph {et~al.}(1998)\citenamefont {Fukuda}
  \emph {et~al.}}]{oscSuperK}%
  \BibitemOpen
  \bibfield  {author} {\bibinfo {author} {\bibfnamefont {Y.}~\bibnamefont
  {Fukuda}} \emph {et~al.} (\bibinfo {collaboration} {Super-Kamiokande}),\
  }\href {\doibase 10.1103/PhysRevLett.81.1562} {\bibfield  {journal} {\bibinfo
   {journal} {Phys. Rev. Lett.}\ }\textbf {\bibinfo {volume} {81}},\ \bibinfo
  {pages} {1562} (\bibinfo {year} {1998})}\BibitemShut {NoStop}%
\bibitem [{\citenamefont {Aharmim}\ \emph {et~al.}(2005)\citenamefont {Aharmim}
  \emph {et~al.}}]{oscSNO}%
  \BibitemOpen
  \bibfield  {author} {\bibinfo {author} {\bibfnamefont {B.}~\bibnamefont
  {Aharmim}} \emph {et~al.} (\bibinfo {collaboration} {SNO}),\ }\href {\doibase
  10.1103/PhysRevC.72.055502} {\bibfield  {journal} {\bibinfo  {journal} {Phys.
  Rev. C}\ }\textbf {\bibinfo {volume} {72}},\ \bibinfo {pages} {055502}
  (\bibinfo {year} {2005})}\BibitemShut {NoStop}%
\bibitem [{\citenamefont {Araki}\ \emph {et~al.}(2005)\citenamefont {Araki}
  \emph {et~al.}}]{oscKamland}%
  \BibitemOpen
  \bibfield  {author} {\bibinfo {author} {\bibfnamefont {T.}~\bibnamefont
  {Araki}} \emph {et~al.} (\bibinfo {collaboration} {KamLAND}),\ }\href
  {\doibase 10.1103/PhysRevLett.94.081801} {\bibfield  {journal} {\bibinfo
  {journal} {Phys. Rev. Lett.}\ }\textbf {\bibinfo {volume} {94}},\ \bibinfo
  {pages} {081801} (\bibinfo {year} {2005})}\BibitemShut {NoStop}%
\bibitem [{\citenamefont {Ahn}\ \emph {et~al.}(2006)\citenamefont {Ahn} \emph
  {et~al.}}]{oscK2K}%
  \BibitemOpen
  \bibfield  {author} {\bibinfo {author} {\bibfnamefont {M.~H.}\ \bibnamefont
  {Ahn}} \emph {et~al.} (\bibinfo {collaboration} {K2K}),\ }\href {\doibase
  10.1103/PhysRevD.74.072003} {\bibfield  {journal} {\bibinfo  {journal} {Phys.
  Rev. D}\ }\textbf {\bibinfo {volume} {74}},\ \bibinfo {pages} {072003}
  (\bibinfo {year} {2006})}\BibitemShut {NoStop}%
\bibitem [{\citenamefont {An}\ \emph {et~al.}(2012)\citenamefont {An} \emph
  {et~al.}}]{oscDayaBay}%
  \BibitemOpen
  \bibfield  {author} {\bibinfo {author} {\bibfnamefont {F.~P.}\ \bibnamefont
  {An}} \emph {et~al.} (\bibinfo {collaboration} {Daya Bay}),\ }\href {\doibase
  10.1103/PhysRevLett.108.171803} {\bibfield  {journal} {\bibinfo  {journal}
  {Phys. Rev. Lett.}\ }\textbf {\bibinfo {volume} {108}},\ \bibinfo {pages}
  {171803} (\bibinfo {year} {2012})}\BibitemShut {NoStop}%
\bibitem [{\citenamefont {Adamson}\ \emph {et~al.}(2014)\citenamefont {Adamson}
  \emph {et~al.}}]{MINOSFinal3Flav}%
  \BibitemOpen
  \bibfield  {author} {\bibinfo {author} {\bibfnamefont {P.}~\bibnamefont
  {Adamson}} \emph {et~al.} (\bibinfo {collaboration} {MINOS}),\ }\href
  {\doibase 10.1103/PhysRevLett.112.191801} {\bibfield  {journal} {\bibinfo
  {journal} {Phys. Rev. Lett.}\ }\textbf {\bibinfo {volume} {112}},\ \bibinfo
  {pages} {191801} (\bibinfo {year} {2014})}\BibitemShut {NoStop}%
\bibitem [{\citenamefont {Schael}\ \emph {et~al.}(2006)\citenamefont {Schael}
  \emph {et~al.}}]{LEP2006}%
  \BibitemOpen
  \bibfield  {author} {\bibinfo {author} {\bibfnamefont {S.}~\bibnamefont
  {Schael}} \emph {et~al.} (\bibinfo {collaboration} {SLD Electroweak Group,
  DELPHI, ALEPH, SLD, SLD Heavy Flavour Group, OPAL, LEP Electroweak Working
  Group, L3}),\ }\href {\doibase 10.1016/j.physrep.2005.12.006} {\bibfield
  {journal} {\bibinfo  {journal} {Phys. Rept.}\ }\textbf {\bibinfo {volume}
  {427}},\ \bibinfo {pages} {257} (\bibinfo {year} {2006})}\BibitemShut
  {NoStop}%
\bibitem [{\citenamefont {Pontecorvo}(1968)}]{Pontecorvo1967}%
  \BibitemOpen
  \bibfield  {author} {\bibinfo {author} {\bibfnamefont {B.}~\bibnamefont
  {Pontecorvo}},\ }\href@noop {} {\bibfield  {journal} {\bibinfo  {journal}
  {Sov. Phys. JETP}\ }\textbf {\bibinfo {volume} {26}},\ \bibinfo {pages} {984}
  (\bibinfo {year} {1968})},\ \bibinfo {note} {[Zh. Eksp. Teor. Fiz. 53, 1717
  (1967)]}\BibitemShut {NoStop}%
\bibitem [{\citenamefont {Gribov}\ and\ \citenamefont
  {Pontecorvo}(1969)}]{Pontecorvo1968}%
  \BibitemOpen
  \bibfield  {author} {\bibinfo {author} {\bibfnamefont {V.~N.}\ \bibnamefont
  {Gribov}}\ and\ \bibinfo {author} {\bibfnamefont {B.}~\bibnamefont
  {Pontecorvo}},\ }\href {\doibase 10.1016/0370-2693(69)90525-5} {\bibfield
  {journal} {\bibinfo  {journal} {Phys. Lett.}\ }\textbf {\bibinfo {volume}
  {28B}},\ \bibinfo {pages} {493} (\bibinfo {year} {1969})}\BibitemShut
  {NoStop}%
\bibitem [{\citenamefont {Maki}\ \emph {et~al.}(1962)\citenamefont {Maki},
  \citenamefont {Nakagawa},\ and\ \citenamefont {Sakata}}]{MNS1962}%
  \BibitemOpen
  \bibfield  {author} {\bibinfo {author} {\bibfnamefont {Z.}~\bibnamefont
  {Maki}}, \bibinfo {author} {\bibfnamefont {M.}~\bibnamefont {Nakagawa}}, \
  and\ \bibinfo {author} {\bibfnamefont {S.}~\bibnamefont {Sakata}},\ }\href
  {\doibase 10.1143/PTP.28.870} {\bibfield  {journal} {\bibinfo  {journal}
  {Prog. Theor. Phys.}\ }\textbf {\bibinfo {volume} {28}},\ \bibinfo {pages}
  {870} (\bibinfo {year} {1962})}\BibitemShut {NoStop}%
\bibitem [{\citenamefont {Aguilar-Arevalo}\ \emph {et~al.}(2001)\citenamefont
  {Aguilar-Arevalo} \emph {et~al.}}]{LSNDSterile}%
  \BibitemOpen
  \bibfield  {author} {\bibinfo {author} {\bibfnamefont {A.}~\bibnamefont
  {Aguilar-Arevalo}} \emph {et~al.} (\bibinfo {collaboration} {LSND}),\ }\href
  {\doibase 10.1103/PhysRevD.64.112007} {\bibfield  {journal} {\bibinfo
  {journal} {Phys. Rev. D}\ }\textbf {\bibinfo {volume} {64}},\ \bibinfo
  {pages} {112007} (\bibinfo {year} {2001})}\BibitemShut {NoStop}%
\bibitem [{\citenamefont {Aguilar-Arevalo}\ \emph {et~al.}(2018)\citenamefont
  {Aguilar-Arevalo} \emph {et~al.}}]{newMiniBooNE}%
  \BibitemOpen
  \bibfield  {author} {\bibinfo {author} {\bibfnamefont {A.}~\bibnamefont
  {Aguilar-Arevalo}} \emph {et~al.} (\bibinfo {collaboration} {MiniBooNE}),\
  }\href {\doibase 10.1103/PhysRevLett.121.221801} {\bibfield  {journal}
  {\bibinfo  {journal} {Phys. Rev. Lett.}\ }\textbf {\bibinfo {volume} {121}},\
  \bibinfo {pages} {221801} (\bibinfo {year} {2018})}\BibitemShut {NoStop}%
\bibitem [{\citenamefont {Acero}\ \emph {et~al.}(2008)\citenamefont {Acero},
  \citenamefont {Giunti},\ and\ \citenamefont {Laveder}}]{GalliumSummary}%
  \BibitemOpen
  \bibfield  {author} {\bibinfo {author} {\bibfnamefont {M.~A.}\ \bibnamefont
  {Acero}}, \bibinfo {author} {\bibfnamefont {C.}~\bibnamefont {Giunti}}, \
  and\ \bibinfo {author} {\bibfnamefont {M.}~\bibnamefont {Laveder}},\ }\href
  {\doibase 10.1103/PhysRevD.78.073009} {\bibfield  {journal} {\bibinfo
  {journal} {Phys. Rev. D}\ }\textbf {\bibinfo {volume} {78}},\ \bibinfo
  {pages} {073009} (\bibinfo {year} {2008})}\BibitemShut {NoStop}%
\bibitem [{\citenamefont {Mention}\ \emph {et~al.}(2011)\citenamefont {Mention}
  \emph {et~al.}}]{ReactorSummary}%
  \BibitemOpen
  \bibfield  {author} {\bibinfo {author} {\bibfnamefont {G.}~\bibnamefont
  {Mention}} \emph {et~al.},\ }\href {\doibase 10.1103/PhysRevD.83.073006}
  {\bibfield  {journal} {\bibinfo  {journal} {Phys. Rev. D}\ }\textbf {\bibinfo
  {volume} {83}},\ \bibinfo {pages} {073006} (\bibinfo {year}
  {2011})}\BibitemShut {NoStop}%
\bibitem [{\citenamefont {An}\ \emph {et~al.}(2017)\citenamefont {An} \emph
  {et~al.}}]{dayaBayFlux}%
  \BibitemOpen
  \bibfield  {author} {\bibinfo {author} {\bibfnamefont {F.~P.}\ \bibnamefont
  {An}} \emph {et~al.} (\bibinfo {collaboration} {Daya Bay}),\ }\href {\doibase
  10.1103/PhysRevLett.118.251801} {\bibfield  {journal} {\bibinfo  {journal}
  {Phys. Rev. Lett.}\ }\textbf {\bibinfo {volume} {118}},\ \bibinfo {pages}
  {251801} (\bibinfo {year} {2017})}\BibitemShut {NoStop}%
\bibitem [{\citenamefont {Huber}(2017)}]{PhysRevLett.118.042502}%
  \BibitemOpen
  \bibfield  {author} {\bibinfo {author} {\bibfnamefont {P.}~\bibnamefont
  {Huber}},\ }\href@noop {} {\bibfield  {journal} {\bibinfo  {journal} {Phys.
  Rev. Lett.}\ }\textbf {\bibinfo {volume} {118}},\ \bibinfo {pages} {042502}
  (\bibinfo {year} {2017})}\BibitemShut {NoStop}%
\bibitem [{\citenamefont {Adamson}\ \emph
  {et~al.}(2016{\natexlab{a}})\citenamefont {Adamson} \emph
  {et~al.}}]{MINOSSterile2016}%
  \BibitemOpen
  \bibfield  {author} {\bibinfo {author} {\bibfnamefont {P.}~\bibnamefont
  {Adamson}} \emph {et~al.} (\bibinfo {collaboration} {MINOS}),\ }\href
  {\doibase 10.1103/PhysRevLett.117.151803} {\bibfield  {journal} {\bibinfo
  {journal} {Phys. Rev. Lett.}\ }\textbf {\bibinfo {volume} {117}},\ \bibinfo
  {pages} {151803} (\bibinfo {year} {2016}{\natexlab{a}})}\BibitemShut
  {NoStop}%
\bibitem [{\citenamefont {Adamson}\ \emph
  {et~al.}(2016{\natexlab{b}})\citenamefont {Adamson} \emph
  {et~al.}}]{MINOSDayaBay2016}%
  \BibitemOpen
  \bibfield  {author} {\bibinfo {author} {\bibfnamefont {P.}~\bibnamefont
  {Adamson}} \emph {et~al.} (\bibinfo {collaboration} {MINOS, Daya Bay}),\
  }\href {\doibase 10.1103/PhysRevLett.117.151801,
  10.1103/PhysRevLett.117.209901} {\bibfield  {journal} {\bibinfo  {journal}
  {Phys. Rev. Lett.}\ }\textbf {\bibinfo {volume} {117}},\ \bibinfo {pages}
  {151801} (\bibinfo {year} {2016}{\natexlab{b}})},\ \bibinfo {note}
  {[Addendum: Phys. Rev. Lett. {\bf 117}, 209901 (2016)]}\BibitemShut {NoStop}%
\bibitem [{\citenamefont {Adamson}\ \emph
  {et~al.}(2016{\natexlab{c}})\citenamefont {Adamson} \emph
  {et~al.}}]{NuMIBeam2016}%
  \BibitemOpen
  \bibfield  {author} {\bibinfo {author} {\bibfnamefont {P.}~\bibnamefont
  {Adamson}} \emph {et~al.},\ }\href {\doibase 10.1016/j.nima.2015.08.063}
  {\bibfield  {journal} {\bibinfo  {journal} {Nucl. Instrum. Meth.}\ }\textbf
  {\bibinfo {volume} {A806}},\ \bibinfo {pages} {279} (\bibinfo {year}
  {2016}{\natexlab{c}})}\BibitemShut {NoStop}%
\bibitem [{\citenamefont {Aartsen}\ \emph {et~al.}(2016)\citenamefont {Aartsen}
  \emph {et~al.}}]{IceCubeSterile}%
  \BibitemOpen
  \bibfield  {author} {\bibinfo {author} {\bibfnamefont {M.~G.}\ \bibnamefont
  {Aartsen}} \emph {et~al.} (\bibinfo {collaboration} {IceCube}),\ }\href
  {\doibase 10.1103/PhysRevLett.117.071801} {\bibfield  {journal} {\bibinfo
  {journal} {Phys. Rev. Lett.}\ }\textbf {\bibinfo {volume} {117}},\ \bibinfo
  {pages} {071801} (\bibinfo {year} {2016})}\BibitemShut {NoStop}%
\bibitem [{\citenamefont {Aartsen}\ \emph {et~al.}(2017)\citenamefont {Aartsen}
  \emph {et~al.}}]{IceCubeDC}%
  \BibitemOpen
  \bibfield  {author} {\bibinfo {author} {\bibfnamefont {M.~G.}\ \bibnamefont
  {Aartsen}} \emph {et~al.} (\bibinfo {collaboration} {IceCube}),\ }\href
  {\doibase 10.1103/PhysRevD.95.112002} {\bibfield  {journal} {\bibinfo
  {journal} {Phys. Rev. D}\ }\textbf {\bibinfo {volume} {95}},\ \bibinfo
  {pages} {112002} (\bibinfo {year} {2017})}\BibitemShut {NoStop}%
\bibitem [{\citenamefont {Michael}\ \emph {et~al.}(2008)\citenamefont {Michael}
  \emph {et~al.}}]{MINOSNIM}%
  \BibitemOpen
  \bibfield  {author} {\bibinfo {author} {\bibfnamefont {D.~G.}\ \bibnamefont
  {Michael}} \emph {et~al.} (\bibinfo {collaboration} {MINOS}),\ }\href
  {\doibase 10.1016/j.nima.2008.08.003} {\bibfield  {journal} {\bibinfo
  {journal} {Nucl. Instrum. Meth.}\ }\textbf {\bibinfo {volume} {A596}},\
  \bibinfo {pages} {190} (\bibinfo {year} {2008})}\BibitemShut {NoStop}%
\bibitem [{\citenamefont {Palazzo}(2013)}]{theta14ZeroJustification}%
  \BibitemOpen
  \bibfield  {author} {\bibinfo {author} {\bibfnamefont {A.}~\bibnamefont
  {Palazzo}},\ }\href {\doibase 10.1142/S0217732313300048} {\bibfield
  {journal} {\bibinfo  {journal} {Mod. Phys. Lett.}\ }\textbf {\bibinfo
  {volume} {A28}},\ \bibinfo {pages} {1330004} (\bibinfo {year}
  {2013})}\BibitemShut {NoStop}%
\bibitem [{\citenamefont {Adamson}\ \emph {et~al.}(2013)\citenamefont {Adamson}
  \emph {et~al.}}]{MINOSDis2013}%
  \BibitemOpen
  \bibfield  {author} {\bibinfo {author} {\bibfnamefont {P.}~\bibnamefont
  {Adamson}} \emph {et~al.} (\bibinfo {collaboration} {MINOS}),\ }\href
  {\doibase 10.1103/PhysRevLett.110.251801} {\bibfield  {journal} {\bibinfo
  {journal} {Phys. Rev. Lett.}\ }\textbf {\bibinfo {volume} {110}},\ \bibinfo
  {pages} {251801} (\bibinfo {year} {2013})}\BibitemShut {NoStop}%
\bibitem [{\citenamefont {Huang}(2015)}]{junting}%
  \BibitemOpen
  \bibfield  {author} {\bibinfo {author} {\bibfnamefont {J.}~\bibnamefont
  {Huang}},\ }\href
  {http://lss.fnal.gov/archive/thesis/2000/fermilab-thesis-2015-06.pdf} {Ph.D.
  thesis},\ \bibinfo  {school} {U. Texas, Austin} (\bibinfo {year}
  {2015})\BibitemShut {NoStop}%
\bibitem [{\citenamefont {Todd}(2018)}]{jacobTodd}%
  \BibitemOpen
  \bibfield  {author} {\bibinfo {author} {\bibfnamefont {J.}~\bibnamefont
  {Todd}},\ }\href
  {http://lss.fnal.gov/archive/thesis/2000/fermilab-thesis-2018-19.pdf} {Ph.D.
  thesis},\ \bibinfo  {school} {U. Cincinnati} (\bibinfo {year}
  {2018})\BibitemShut {NoStop}%
%%CITATION = FERMILAB-THESIS-2018-19;%%
\bibitem [{Note1()}]{Note1}%
  \BibitemOpen
  \bibinfo {note} {See Supplemental Material for further details, which
  contains Refs.~\cite
  {Asimov,ref:fluka1,ref:fluka2,NA49Exp,MinervaExp,ref:matrixpram,CCFRExp,NOMADExp,ref:flugg,Geant4:2003,Geant4:2006,MEC,MECMiniBooNE,RPA,RPAMinerva,MinosCalDet,intranuke,HadESyst,Conway:2011in,ATLAS_Asimov,katz_uli_2018,Patterson:2007zz,JAMES1975343,CovMx}}\BibitemShut
  {NoStop}%
\bibitem [{\citenamefont {Adamson}\ \emph {et~al.}(2008)\citenamefont {Adamson}
  \emph {et~al.}}]{minosDisPRD}%
  \BibitemOpen
  \bibfield  {author} {\bibinfo {author} {\bibfnamefont {P.}~\bibnamefont
  {Adamson}} \emph {et~al.} (\bibinfo {collaboration} {MINOS}),\ }\href
  {\doibase 10.1103/PhysRevD.77.072002} {\bibfield  {journal} {\bibinfo
  {journal} {Phys. Rev. D}\ }\textbf {\bibinfo {volume} {77}},\ \bibinfo
  {pages} {072002} (\bibinfo {year} {2008})}\BibitemShut {NoStop}%
\bibitem [{\citenamefont {Aliaga}\ \emph {et~al.}(2016)\citenamefont {Aliaga}
  \emph {et~al.}}]{MinervaFlux2016}%
  \BibitemOpen
  \bibfield  {author} {\bibinfo {author} {\bibfnamefont {L.}~\bibnamefont
  {Aliaga}} \emph {et~al.} (\bibinfo {collaboration} {MINERvA}),\ }\href
  {\doibase 10.1103/PhysRevD.94.092005, 10.1103/PhysRevD.95.039903} {\bibfield
  {journal} {\bibinfo  {journal} {Phys. Rev. D}\ }\textbf {\bibinfo {volume}
  {94}},\ \bibinfo {pages} {092005} (\bibinfo {year} {2016})},\ \bibinfo {note}
  {[Addendum: Phys. Rev. D {\bf 95}, 039903 (2017)]}\BibitemShut {NoStop}%
\bibitem [{\citenamefont {Aliaga~Soplin}(2016)}]{leoThesis}%
  \BibitemOpen
  \bibfield  {author} {\bibinfo {author} {\bibfnamefont {L.}~\bibnamefont
  {Aliaga~Soplin}},\ }\href
  {http://lss.fnal.gov/cgi-bin/find_paper.pl?thesis-2016-03} {Ph.D. thesis},\
  \bibinfo  {school} {College of William and Mary} (\bibinfo {year}
  {2016})\BibitemShut {NoStop}%
%%CITATION = FERMILAB-THESIS-2016-03;%%
\bibitem [{Note2()}]{Note2}%
  \BibitemOpen
  \bibinfo {note} {The MINERvA publication~\cite {MinervaFlux2016} incorporates
  the $\nu _e e \rightarrow \nu _e e$ scattering rate in the published fluxes.
  This constraint is not used in this analysis.}\BibitemShut {Stop}%
\bibitem [{\citenamefont {Lebedev}(2007)}]{mippFluka}%
  \BibitemOpen
  \bibfield  {author} {\bibinfo {author} {\bibfnamefont {A.}~\bibnamefont
  {Lebedev}},\ }\href
  {http://lss.fnal.gov/cgi-bin/find_paper.pl?thesis-2007-76} {Ph.D. thesis},\
  \bibinfo  {school} {Harvard University} (\bibinfo {year} {2007})\BibitemShut
  {NoStop}%
%%CITATION = FERMILAB-THESIS-2007-76;%%
\bibitem [{\citenamefont {Timmons}(2016)}]{ashleyThesis}%
  \BibitemOpen
  \bibfield  {author} {\bibinfo {author} {\bibfnamefont {A.~M.}\ \bibnamefont
  {Timmons}},\ }\href@noop {} {Ph.D. thesis},\ \bibinfo  {school} {University
  of Manchester} (\bibinfo {year} {2016})\BibitemShut {NoStop}%
\bibitem [{\citenamefont {Adamson}\ \emph {et~al.}(2011)\citenamefont {Adamson}
  \emph {et~al.}}]{minosSysts}%
  \BibitemOpen
  \bibfield  {author} {\bibinfo {author} {\bibfnamefont {P.}~\bibnamefont
  {Adamson}} \emph {et~al.} (\bibinfo {collaboration} {MINOS}),\ }\href
  {\doibase 10.1103/PhysRevLett.106.181801} {\bibfield  {journal} {\bibinfo
  {journal} {Phys. Rev. Lett.}\ }\textbf {\bibinfo {volume} {106}},\ \bibinfo
  {pages} {181801} (\bibinfo {year} {2011})}\BibitemShut {NoStop}%
\bibitem [{\citenamefont {Gallagher}(2002)}]{NEUGEN}%
  \BibitemOpen
  \bibfield  {author} {\bibinfo {author} {\bibfnamefont {H.}~\bibnamefont
  {Gallagher}},\ }\href {\doibase 10.1016/S0920-5632(02)01775-9} {\bibfield
  {journal} {\bibinfo  {journal} {Nucl. Phys. Proc. Suppl.}\ }\textbf {\bibinfo
  {volume} {112}},\ \bibinfo {pages} {188} (\bibinfo {year}
  {2002})}\BibitemShut {NoStop}%
\bibitem [{\citenamefont {Auchincloss}\ \emph {et~al.}(1990)\citenamefont
  {Auchincloss} \emph {et~al.}}]{CCFRXSec}%
  \BibitemOpen
  \bibfield  {author} {\bibinfo {author} {\bibfnamefont {P.~S.}\ \bibnamefont
  {Auchincloss}} \emph {et~al.} (\bibinfo {collaboration} {CCFR}),\ }\href@noop
  {} {\bibfield  {journal} {\bibinfo  {journal} {Z Phys C}\ }\textbf {\bibinfo
  {volume} {48}},\ \bibinfo {pages} {411} (\bibinfo {year} {1990})}\BibitemShut
  {NoStop}%
\bibitem [{\citenamefont {Koba}\ \emph {et~al.}(1972)\citenamefont {Koba},
  \citenamefont {Nielsen},\ and\ \citenamefont {Olesen}}]{KNO}%
  \BibitemOpen
  \bibfield  {author} {\bibinfo {author} {\bibfnamefont {Z.}~\bibnamefont
  {Koba}}, \bibinfo {author} {\bibfnamefont {H.}~\bibnamefont {Nielsen}}, \
  and\ \bibinfo {author} {\bibfnamefont {P.}~\bibnamefont {Olesen}},\ }\href
  {\doibase https://doi.org/10.1016/0550-3213(72)90551-2} {\bibfield  {journal}
  {\bibinfo  {journal} {Nucl. Phys. B}\ }\textbf {\bibinfo {volume} {40}},\
  \bibinfo {pages} {317 } (\bibinfo {year} {1972})}\BibitemShut {NoStop}%
\bibitem [{\citenamefont {Zwaska}\ \emph {et~al.}(2006)\citenamefont {Zwaska}
  \emph {et~al.}}]{beamSyst}%
  \BibitemOpen
  \bibfield  {author} {\bibinfo {author} {\bibfnamefont {R.}~\bibnamefont
  {Zwaska}} \emph {et~al.},\ }\href {\doibase 10.1016/j.nima.2006.08.031}
  {\bibfield  {journal} {\bibinfo  {journal} {Nucl. Instrum. Meth.}\ }\textbf
  {\bibinfo {volume} {A568}},\ \bibinfo {pages} {548} (\bibinfo {year}
  {2006})}\BibitemShut {NoStop}%
\bibitem [{\citenamefont {Pavlovic}(2008)}]{Zarko}%
  \BibitemOpen
  \bibfield  {author} {\bibinfo {author} {\bibfnamefont {Z.}~\bibnamefont
  {Pavlovic}},\ }\href
  {http://lss.fnal.gov/archive/thesis/2000/fermilab-thesis-2008-59.pdf} {Ph.D.
  thesis},\ \bibinfo  {school} {U. Texas, Austin} (\bibinfo {year}
  {2008})\BibitemShut {NoStop}%
%%CITATION = FERMILAB-THESIS-2008-59;%%
\bibitem [{\citenamefont {Devan}(2015)}]{aDevan}%
  \BibitemOpen
  \bibfield  {author} {\bibinfo {author} {\bibfnamefont {A.~V.}\ \bibnamefont
  {Devan}},\ }\href
  {http://lss.fnal.gov/archive/thesis/2000/fermilab-thesis-2015-12.pdf} {Ph.D.
  thesis},\ \bibinfo  {school} {College of William and Mary} (\bibinfo {year}
  {2015})\BibitemShut {NoStop}%
%%CITATION = FERMILAB-THESIS-2015-12;%%
\bibitem [{\citenamefont {Fogli}\ \emph {et~al.}(2012)\citenamefont {Fogli}
  \emph {et~al.}}]{solarParsJustification}%
  \BibitemOpen
  \bibfield  {author} {\bibinfo {author} {\bibfnamefont {G.~L.}\ \bibnamefont
  {Fogli}} \emph {et~al.},\ }\href {\doibase 10.1103/PhysRevD.86.013012}
  {\bibfield  {journal} {\bibinfo  {journal} {Phys. Rev. D}\ }\textbf {\bibinfo
  {volume} {86}},\ \bibinfo {pages} {013012} (\bibinfo {year}
  {2012})}\BibitemShut {NoStop}%
\bibitem [{\citenamefont {Parke}\ and\ \citenamefont
  {Ross-Lonergan}(2016)}]{PhysRevD.93.113009}%
  \BibitemOpen
  \bibfield  {author} {\bibinfo {author} {\bibfnamefont {S.}~\bibnamefont
  {Parke}}\ and\ \bibinfo {author} {\bibfnamefont {M.}~\bibnamefont
  {Ross-Lonergan}},\ }\href {\doibase 10.1103/PhysRevD.93.113009} {\bibfield
  {journal} {\bibinfo  {journal} {Phys. Rev. D}\ }\textbf {\bibinfo {volume}
  {93}},\ \bibinfo {pages} {113009} (\bibinfo {year} {2016})}\BibitemShut
  {NoStop}%
\bibitem [{\citenamefont {Feldman}\ and\ \citenamefont
  {Cousins}(1998)}]{fcPaper}%
  \BibitemOpen
  \bibfield  {author} {\bibinfo {author} {\bibfnamefont {G.~J.}\ \bibnamefont
  {Feldman}}\ and\ \bibinfo {author} {\bibfnamefont {R.~D.}\ \bibnamefont
  {Cousins}},\ }\href {\doibase 10.1103/PhysRevD.57.3873} {\bibfield  {journal}
  {\bibinfo  {journal} {Phys. Rev. D}\ }\textbf {\bibinfo {volume} {57}},\
  \bibinfo {pages} {3873} (\bibinfo {year} {1998})}\BibitemShut {NoStop}%
\bibitem [{\citenamefont {Esteban}\ \emph {et~al.}(2017)\citenamefont
  {Esteban}, \citenamefont {Gonzalez-Garcia}, \citenamefont {Maltoni},
  \citenamefont {Martinez-Soler},\ and\ \citenamefont {Schwetz}}]{NuFit}%
  \BibitemOpen
  \bibfield  {author} {\bibinfo {author} {\bibfnamefont {I.}~\bibnamefont
  {Esteban}}, \bibinfo {author} {\bibfnamefont {M.~C.}\ \bibnamefont
  {Gonzalez-Garcia}}, \bibinfo {author} {\bibfnamefont {M.}~\bibnamefont
  {Maltoni}}, \bibinfo {author} {\bibfnamefont {I.}~\bibnamefont
  {Martinez-Soler}}, \ and\ \bibinfo {author} {\bibfnamefont {T.}~\bibnamefont
  {Schwetz}},\ }\href {\doibase 10.1007/JHEP01(2017)087} {\bibfield  {journal}
  {\bibinfo  {journal} {JHEP}\ }\textbf {\bibinfo {volume} {01}},\ \bibinfo
  {pages} {087} (\bibinfo {year} {2017})}\BibitemShut {NoStop}%
\bibitem [{\citenamefont {Abe}\ \emph {et~al.}(2015)\citenamefont {Abe} \emph
  {et~al.}}]{SKSterile}%
  \BibitemOpen
  \bibfield  {author} {\bibinfo {author} {\bibfnamefont {K.}~\bibnamefont
  {Abe}} \emph {et~al.} (\bibinfo {collaboration} {Super-Kamiokande}),\ }\href
  {\doibase 10.1103/PhysRevD.91.052019} {\bibfield  {journal} {\bibinfo
  {journal} {Phys. Rev. D}\ }\textbf {\bibinfo {volume} {91}},\ \bibinfo
  {pages} {052019} (\bibinfo {year} {2015})}\BibitemShut {NoStop}%
\bibitem [{\citenamefont {Dydak}\ \emph {et~al.}(1984)\citenamefont {Dydak}
  \emph {et~al.}}]{CDHSWSterile}%
  \BibitemOpen
  \bibfield  {author} {\bibinfo {author} {\bibfnamefont {F.}~\bibnamefont
  {Dydak}} \emph {et~al.} (\bibinfo {collaboration} {CDHSW}),\ }\href {\doibase
  10.1016/0370-2693(84)90688-9} {\bibfield  {journal} {\bibinfo  {journal}
  {Phys. Lett.}\ }\textbf {\bibinfo {volume} {B134}},\ \bibinfo {pages} {281}
  (\bibinfo {year} {1984})}\BibitemShut {NoStop}%
%%CITATION = PHLTA,B134,281;%%
\bibitem [{\citenamefont {Stockdale}\ \emph {et~al.}(1984)\citenamefont
  {Stockdale} \emph {et~al.}}]{CCFRSterile}%
  \BibitemOpen
  \bibfield  {author} {\bibinfo {author} {\bibfnamefont {I.~E.}\ \bibnamefont
  {Stockdale}} \emph {et~al.} (\bibinfo {collaboration} {CCFR}),\ }\href
  {\doibase 10.1103/PhysRevLett.52.1384} {\bibfield  {journal} {\bibinfo
  {journal} {Phys. Rev. Lett.}\ }\textbf {\bibinfo {volume} {52}},\ \bibinfo
  {pages} {1384} (\bibinfo {year} {1984})}\BibitemShut {NoStop}%
%%CITATION = PRLTA,52,1384;%%
\bibitem [{\citenamefont {Mahn}\ \emph {et~al.}(2012)\citenamefont {Mahn} \emph
  {et~al.}}]{SciMBSterile}%
  \BibitemOpen
  \bibfield  {author} {\bibinfo {author} {\bibfnamefont {K.~B.~M.}\
  \bibnamefont {Mahn}} \emph {et~al.} (\bibinfo {collaboration} {SciBooNE,
  MiniBooNE}),\ }\href {\doibase 10.1103/PhysRevD.85.032007} {\bibfield
  {journal} {\bibinfo  {journal} {Phys. Rev.}\ }\textbf {\bibinfo {volume}
  {D85}},\ \bibinfo {pages} {032007} (\bibinfo {year} {2012})}\BibitemShut
  {NoStop}%
\bibitem [{\citenamefont {Gariazzo}\ \emph {et~al.}(2016)\citenamefont
  {Gariazzo}, \citenamefont {Giunti}, \citenamefont {Laveder}, \citenamefont
  {Li},\ and\ \citenamefont {Zavanin}}]{GiuntiSterile}%
  \BibitemOpen
  \bibfield  {author} {\bibinfo {author} {\bibfnamefont {S.}~\bibnamefont
  {Gariazzo}}, \bibinfo {author} {\bibfnamefont {C.}~\bibnamefont {Giunti}},
  \bibinfo {author} {\bibfnamefont {M.}~\bibnamefont {Laveder}}, \bibinfo
  {author} {\bibfnamefont {Y.~F.}\ \bibnamefont {Li}}, \ and\ \bibinfo {author}
  {\bibfnamefont {E.~M.}\ \bibnamefont {Zavanin}},\ }\href {\doibase
  10.1088/0954-3899/43/3/033001} {\bibfield  {journal} {\bibinfo  {journal} {J.
  Phys.}\ }\textbf {\bibinfo {volume} {G43}},\ \bibinfo {pages} {033001}
  (\bibinfo {year} {2016})}\BibitemShut {NoStop}%
\bibitem [{\citenamefont {Cowan}\ \emph {et~al.}(2011)\citenamefont {Cowan},
  \citenamefont {Cranmer}, \citenamefont {Gross},\ and\ \citenamefont
  {Vitells}}]{Asimov}%
  \BibitemOpen
  \bibfield  {author} {\bibinfo {author} {\bibfnamefont {G.}~\bibnamefont
  {Cowan}}, \bibinfo {author} {\bibfnamefont {K.}~\bibnamefont {Cranmer}},
  \bibinfo {author} {\bibfnamefont {E.}~\bibnamefont {Gross}}, \ and\ \bibinfo
  {author} {\bibfnamefont {O.}~\bibnamefont {Vitells}},\ }\href {\doibase
  10.1140/epjc/s10052-011-1554-0} {\bibfield  {journal} {\bibinfo  {journal}
  {Eur. Phys. J. C}\ }\textbf {\bibinfo {volume} {71}},\ \bibinfo {pages}
  {1554} (\bibinfo {year} {2011})}\BibitemShut {NoStop}%
\bibitem [{\citenamefont {Bohlen}\ \emph {et~al.}(2014)\citenamefont {Bohlen},
  \citenamefont {Cerutti}, \citenamefont {Chin}, \citenamefont {Fasso},
  \citenamefont {Ferrari}, \citenamefont {Ortega}, \citenamefont {Mairiani},
  \citenamefont {Sala}, \citenamefont {Smirnov},\ and\ \citenamefont
  {Vlachoudis}}]{ref:fluka1}%
  \BibitemOpen
  \bibfield  {author} {\bibinfo {author} {\bibfnamefont {T.}~\bibnamefont
  {Bohlen}}, \bibinfo {author} {\bibfnamefont {F.}~\bibnamefont {Cerutti}},
  \bibinfo {author} {\bibfnamefont {M.}~\bibnamefont {Chin}}, \bibinfo {author}
  {\bibfnamefont {A.}~\bibnamefont {Fasso}}, \bibinfo {author} {\bibfnamefont
  {A.}~\bibnamefont {Ferrari}}, \bibinfo {author} {\bibfnamefont
  {P.}~\bibnamefont {Ortega}}, \bibinfo {author} {\bibfnamefont
  {A.}~\bibnamefont {Mairiani}}, \bibinfo {author} {\bibfnamefont {P.~R.}\
  \bibnamefont {Sala}}, \bibinfo {author} {\bibfnamefont {G.}~\bibnamefont
  {Smirnov}}, \ and\ \bibinfo {author} {\bibfnamefont {V.}~\bibnamefont
  {Vlachoudis}},\ }\href@noop {} {\bibfield  {journal} {\bibinfo  {journal}
  {Nucl. Data Sheets}\ }\textbf {\bibinfo {volume} {120}},\ \bibinfo {pages}
  {211} (\bibinfo {year} {2014})}\BibitemShut {NoStop}%
\bibitem [{\citenamefont {Ferrari}\ \emph {et~al.}(2005)\citenamefont
  {Ferrari}, \citenamefont {Sala}, \citenamefont {Fasso},\ and\ \citenamefont
  {Ranft}}]{ref:fluka2}%
  \BibitemOpen
  \bibfield  {author} {\bibinfo {author} {\bibfnamefont {A.}~\bibnamefont
  {Ferrari}}, \bibinfo {author} {\bibfnamefont {P.~R.}\ \bibnamefont {Sala}},
  \bibinfo {author} {\bibfnamefont {A.}~\bibnamefont {Fasso}}, \ and\ \bibinfo
  {author} {\bibfnamefont {J.}~\bibnamefont {Ranft}},\ }\href@noop {}
  {\bibfield  {journal} {\bibinfo  {journal} {CERN Report No. CERN-2005-010,
  2005}\ } (\bibinfo {year} {2005})}\BibitemShut {NoStop}%
\bibitem [{\citenamefont {Alt}\ \emph {et~al.}(2007)\citenamefont {Alt} \emph
  {et~al.}}]{NA49Exp}%
  \BibitemOpen
  \bibfield  {author} {\bibinfo {author} {\bibfnamefont {C.}~\bibnamefont
  {Alt}} \emph {et~al.} (\bibinfo {collaboration} {NA49}),\ }\href@noop {}
  {\bibfield  {journal} {\bibinfo  {journal} {Eur. Phys. J. C}\ }\textbf
  {\bibinfo {volume} {49}},\ \bibinfo {pages} {897} (\bibinfo {year}
  {2007})}\BibitemShut {NoStop}%
\bibitem [{\citenamefont {Aliaga}\ \emph {et~al.}(2014)\citenamefont {Aliaga}
  \emph {et~al.}}]{MinervaExp}%
  \BibitemOpen
  \bibfield  {author} {\bibinfo {author} {\bibfnamefont {L.}~\bibnamefont
  {Aliaga}} \emph {et~al.} (\bibinfo {collaboration} {MINERvA}),\ }\href@noop
  {} {\bibfield  {journal} {\bibinfo  {journal} {Nucl. Instrum. Methods A}\
  }\textbf {\bibinfo {volume} {743}},\ \bibinfo {pages} {130} (\bibinfo {year}
  {2014})}\BibitemShut {NoStop}%
\bibitem [{\citenamefont {Harari}\ and\ \citenamefont
  {Leurer}(1986)}]{ref:matrixpram}%
  \BibitemOpen
  \bibfield  {author} {\bibinfo {author} {\bibfnamefont {H.}~\bibnamefont
  {Harari}}\ and\ \bibinfo {author} {\bibfnamefont {M.}~\bibnamefont
  {Leurer}},\ }\href@noop {} {\bibfield  {journal} {\bibinfo  {journal} {Phys.
  Lett. B}\ }\textbf {\bibinfo {volume} {181}},\ \bibinfo {pages} {123}
  (\bibinfo {year} {1986})}\BibitemShut {NoStop}%
\bibitem [{\citenamefont {Avvakumov}\ \emph {et~al.}(1999)\citenamefont
  {Avvakumov} \emph {et~al.}}]{CCFRExp}%
  \BibitemOpen
  \bibfield  {author} {\bibinfo {author} {\bibfnamefont {S.}~\bibnamefont
  {Avvakumov}} \emph {et~al.} (\bibinfo {collaboration} {CCFR/NuTeV}),\
  }\href@noop {} {\bibfield  {journal} {\bibinfo  {journal} {Nucl. Phys. Proc.
  Suppl.}\ }\textbf {\bibinfo {volume} {78}},\ \bibinfo {pages} {232} (\bibinfo
  {year} {1999})}\BibitemShut {NoStop}%
\bibitem [{\citenamefont {Altegoer}\ \emph {et~al.}(1998)\citenamefont
  {Altegoer} \emph {et~al.}}]{NOMADExp}%
  \BibitemOpen
  \bibfield  {author} {\bibinfo {author} {\bibfnamefont {J.}~\bibnamefont
  {Altegoer}} \emph {et~al.} (\bibinfo {collaboration} {NOMAD}),\ }\href@noop
  {} {\bibfield  {journal} {\bibinfo  {journal} {Nucl. Instrum. Methods A}\
  }\textbf {\bibinfo {volume} {404}} (\bibinfo {year} {1998})}\BibitemShut
  {NoStop}%
\bibitem [{\citenamefont {Campanella}\ \emph {et~al.}(1999)\citenamefont
  {Campanella}, \citenamefont {Ferrari}, \citenamefont {Sala},\ and\
  \citenamefont {Vanini}}]{ref:flugg}%
  \BibitemOpen
  \bibfield  {author} {\bibinfo {author} {\bibfnamefont {M.}~\bibnamefont
  {Campanella}}, \bibinfo {author} {\bibfnamefont {A.}~\bibnamefont {Ferrari}},
  \bibinfo {author} {\bibfnamefont {P.~R.}\ \bibnamefont {Sala}}, \ and\
  \bibinfo {author} {\bibfnamefont {S.}~\bibnamefont {Vanini}},\ }\href@noop {}
  {\bibfield  {journal} {\bibinfo  {journal} {CERN Report No.
  CERN-ATL-SOFT-99-004, 1999}\ } (\bibinfo {year} {1999})}\BibitemShut
  {NoStop}%
\bibitem [{\citenamefont {Agostinelli}\ \emph {et~al.}(2003)\citenamefont
  {Agostinelli} \emph {et~al.}}]{Geant4:2003}%
  \BibitemOpen
  \bibfield  {author} {\bibinfo {author} {\bibfnamefont {S.}~\bibnamefont
  {Agostinelli}} \emph {et~al.},\ }\href@noop {} {\bibfield  {journal}
  {\bibinfo  {journal} {Nucl. Instrum. Meth.}\ }\textbf {\bibinfo {volume}
  {A506}},\ \bibinfo {pages} {250} (\bibinfo {year} {2003})}\BibitemShut
  {NoStop}%
\bibitem [{\citenamefont {Allison}\ \emph {et~al.}(2006)\citenamefont {Allison}
  \emph {et~al.}}]{Geant4:2006}%
  \BibitemOpen
  \bibfield  {author} {\bibinfo {author} {\bibfnamefont {J.}~\bibnamefont
  {Allison}} \emph {et~al.},\ }\href@noop {} {\bibfield  {journal} {\bibinfo
  {journal} {IEEE Trans. Nucl. Sci.}\ }\textbf {\bibinfo {volume} {53}},\
  \bibinfo {pages} {270} (\bibinfo {year} {2006})}\BibitemShut {NoStop}%
\bibitem [{\citenamefont {Katori}(2015)}]{MEC}%
  \BibitemOpen
  \bibfield  {author} {\bibinfo {author} {\bibfnamefont {T.}~\bibnamefont
  {Katori}},\ }\href {https://aip.scitation.org/doi/abs/10.1063/1.4919465}
  {\bibfield  {journal} {\bibinfo  {journal} {AIP Conf. Proc.}\ }\textbf
  {\bibinfo {volume} {1663}},\ \bibinfo {pages} {030001} (\bibinfo {year}
  {2015})}\BibitemShut {NoStop}%
\bibitem [{\citenamefont {Aguilar-Arevalo}\ \emph {et~al.}(2010)\citenamefont
  {Aguilar-Arevalo} \emph {et~al.}}]{MECMiniBooNE}%
  \BibitemOpen
  \bibfield  {author} {\bibinfo {author} {\bibfnamefont {A.~A.}\ \bibnamefont
  {Aguilar-Arevalo}} \emph {et~al.} (\bibinfo {collaboration} {MiniBooNE}),\
  }\href@noop {} {\bibfield  {journal} {\bibinfo  {journal} {Phys. Rev. D}\
  }\textbf {\bibinfo {volume} {81}},\ \bibinfo {pages} {092005} (\bibinfo
  {year} {2010})}\BibitemShut {NoStop}%
\bibitem [{\citenamefont {Nieves}\ \emph {et~al.}(2004)\citenamefont {Nieves},
  \citenamefont {Amaro},\ and\ \citenamefont {Valverde}}]{RPA}%
  \BibitemOpen
  \bibfield  {author} {\bibinfo {author} {\bibfnamefont {J.}~\bibnamefont
  {Nieves}}, \bibinfo {author} {\bibfnamefont {J.~E.}\ \bibnamefont {Amaro}}, \
  and\ \bibinfo {author} {\bibfnamefont {M.}~\bibnamefont {Valverde}},\
  }\href@noop {} {\bibfield  {journal} {\bibinfo  {journal} {Phys. Rev. C}\
  }\textbf {\bibinfo {volume} {70}},\ \bibinfo {pages} {055503} (\bibinfo
  {year} {2004})},\ \bibinfo {note} {; {\bf 72}, 019902 (2005)
  [Erratum]}\BibitemShut {NoStop}%
\bibitem [{\citenamefont {Rodrigues}\ \emph {et~al.}(2016)\citenamefont
  {Rodrigues} \emph {et~al.}}]{RPAMinerva}%
  \BibitemOpen
  \bibfield  {author} {\bibinfo {author} {\bibfnamefont {P.~A.}\ \bibnamefont
  {Rodrigues}} \emph {et~al.} (\bibinfo {collaboration} {MINERvA}),\
  }\href@noop {} {\bibfield  {journal} {\bibinfo  {journal} {Phys. Rev. Lett.}\
  }\textbf {\bibinfo {volume} {116}},\ \bibinfo {pages} {071802} (\bibinfo
  {year} {2016})}\BibitemShut {NoStop}%
\bibitem [{\citenamefont {Cabrera}\ \emph {et~al.}(2009)\citenamefont {Cabrera}
  \emph {et~al.}}]{MinosCalDet}%
  \BibitemOpen
  \bibfield  {author} {\bibinfo {author} {\bibfnamefont {A.}~\bibnamefont
  {Cabrera}} \emph {et~al.} (\bibinfo {collaboration} {MINOS}),\ }\href@noop {}
  {\bibfield  {journal} {\bibinfo  {journal} {Nucl. Instrum. Methods A}\
  }\textbf {\bibinfo {volume} {609}},\ \bibinfo {pages} {106} (\bibinfo {year}
  {2009})}\BibitemShut {NoStop}%
\bibitem [{\citenamefont {Merenyi}(1990)}]{intranuke}%
  \BibitemOpen
  \bibfield  {author} {\bibinfo {author} {\bibfnamefont {R.}~\bibnamefont
  {Merenyi}},\ }\href@noop {} {Ph.D. thesis},\ \bibinfo  {school} {Tufts
  University} (\bibinfo {year} {1990})\BibitemShut {NoStop}%
\bibitem [{\citenamefont {Dytman}\ \emph {et~al.}(2008)\citenamefont {Dytman},
  \citenamefont {Gallagher},\ and\ \citenamefont {Kordosky}}]{HadESyst}%
  \BibitemOpen
  \bibfield  {author} {\bibinfo {author} {\bibfnamefont {S.}~\bibnamefont
  {Dytman}}, \bibinfo {author} {\bibfnamefont {H.}~\bibnamefont {Gallagher}}, \
  and\ \bibinfo {author} {\bibfnamefont {M.}~\bibnamefont {Kordosky}},\
  }\href@noop {} {\bibfield  {journal} {\bibinfo  {journal} {arXiv:0806.2119}\
  } (\bibinfo {year} {2008})}\BibitemShut {NoStop}%
\bibitem [{\citenamefont {Conway}(2011)}]{Conway:2011in}%
  \BibitemOpen
  \bibfield  {author} {\bibinfo {author} {\bibfnamefont {J.~S.}\ \bibnamefont
  {Conway}},\ }in\ \href {\doibase 10.5170/CERN-2011-006.115} {\emph {\bibinfo
  {booktitle} {{Proceedings, PHYSTAT 2011 Workshop on Statistical Issues
  Related to Discovery Claims in Search Experiments and Unfolding, CERN,Geneva,
  Switzerland 17-20 January 2011}}}}\ (\bibinfo {year} {2011})\ pp.\ \bibinfo
  {pages} {115--120},\ \Eprint {http://arxiv.org/abs/1103.0354}
  {arXiv:1103.0354 [physics.data-an]} \BibitemShut {NoStop}%
%%CITATION = ARXIV:1103.0354;%%
\bibitem [{\citenamefont {Berger}(2018)}]{ATLAS_Asimov}%
  \BibitemOpen
  \bibfield  {author} {\bibinfo {author} {\bibfnamefont {N.}~\bibnamefont
  {Berger}},\ }\href@noop {} {}\bibinfo {howpublished} {{(private
  communication)}} (\bibinfo {year} {2018})\BibitemShut {NoStop}%
\bibitem [{\citenamefont {Katz}(2018)}]{katz_uli_2018}%
  \BibitemOpen
  \bibfield  {author} {\bibinfo {author} {\bibfnamefont {U.}~\bibnamefont
  {Katz}},\ }\href {\doibase 10.5281/zenodo.1287686} {\enquote {\bibinfo
  {title} {Future neutrino telescopes in water and ice},}\ } (\bibinfo {year}
  {2018})\BibitemShut {NoStop}%
\bibitem [{\citenamefont {Patterson}(2007)}]{Patterson:2007zz}%
  \BibitemOpen
  \bibfield  {author} {\bibinfo {author} {\bibfnamefont {R.~B.}\ \bibnamefont
  {Patterson}},\ }\emph {\bibinfo {title} {{A Search for Muon Neutrino to
  Electron Neutrino Oscillations at $\delta(m^2) >0.1$ eV$^2$}}},\ \href
  {\doibase 10.2172/917855} {Ph.D. thesis},\ \bibinfo  {school} {Princeton U.}
  (\bibinfo {year} {2007})\BibitemShut {NoStop}%
%%CITATION = FERMILAB-THESIS-2007-19;%%
\bibitem [{\citenamefont {James}\ and\ \citenamefont
  {Roos}(1975)}]{JAMES1975343}%
  \BibitemOpen
  \bibfield  {author} {\bibinfo {author} {\bibfnamefont {F.}~\bibnamefont
  {James}}\ and\ \bibinfo {author} {\bibfnamefont {M.}~\bibnamefont {Roos}},\
  }\href {\doibase https://doi.org/10.1016/0010-4655(75)90039-9} {\bibfield
  {journal} {\bibinfo  {journal} {Computer Physics Communications}\ }\textbf
  {\bibinfo {volume} {10}},\ \bibinfo {pages} {343 } (\bibinfo {year}
  {1975})}\BibitemShut {NoStop}%
\bibitem [{\citenamefont {Eaton}(1983)}]{CovMx}%
  \BibitemOpen
  \bibfield  {author} {\bibinfo {author} {\bibfnamefont {M.~L.}\ \bibnamefont
  {Eaton}},\ }\href@noop {} {\emph {\bibinfo {title} {Multivariate Statistics:
  a Vector Space Approach}}}\ (\bibinfo  {publisher} {John Wiley and Sons},\
  \bibinfo {year} {1983})\BibitemShut {NoStop}%
\end{thebibliography}%

\end{document}